\documentclass[aps,pra,reprint,showpacs,nofootinbib,superscriptaddress]{revtex4-1}
\usepackage{graphicx, comment}
\usepackage{amsfonts}
\usepackage{amsmath,amssymb}
\usepackage{bm}
\usepackage{color}
\usepackage{epstopdf}
\usepackage{epsfig}
\usepackage{verbatim}
\usepackage{lineno}
\usepackage{subcaption}
\usepackage[thicklines]{cancel}
\usepackage{url}
\usepackage{soul}

\usepackage{tikz}
\usepackage{pgf-umlcd}
\usepackage[edges]{forest}
\usetikzlibrary{arrows.meta,shadows.blur}
\colorlet{myyellow}{yellow!50}
\usepackage{float}
\usepackage[ruled,vlined]{algorithm2e}
\usepackage{dblfloatfix}
\usepackage{graphicx,float}
\usetikzlibrary{shapes,arrows}
\usepackage{amsmath,bm,times}
\usepackage{pgfplots}
\usepackage{pgfplotstable}
\usepackage{subcaption}
\usepackage[export]{adjustbox}
\usepackage{bera}
\usepackage{listings,mdframed}
\usepackage{xcolor}

\usepackage{pythonhighlight}
\definecolor{comment-text-color}{rgb}{0,0.8,0.6}
\lstdefinelanguage{json}{
    basicstyle=\normalfont\ttfamily\footnotesize,
    numbers=left,
    numberstyle=\scriptsize,
    stepnumber=1,
    numbersep=8pt,
    showstringspaces=false,
    breaklines=true,
    frame=lines,
    backgroundcolor=\color{background},
    literate=
     *{0}{{{\color{numb}0}}}{1}
      {1}{{{\color{numb}1}}}{1}
      {2}{{{\color{numb}2}}}{1}
      {3}{{{\color{numb}3}}}{1}
      {4}{{{\color{numb}4}}}{1}
      {5}{{{\color{numb}5}}}{1}
      {6}{{{\color{numb}6}}}{1}
      {7}{{{\color{numb}7}}}{1}
      {8}{{{\color{numb}8}}}{1}
      {9}{{{\color{numb}9}}}{1}
      {:}{{{\color{punct}{:}}}}{1}
      {,}{{{\color{punct}{,}}}}{1}
      {\{}{{{\color{delim}{\{}}}}{1}
      {\}}{{{\color{delim}{\}}}}}{1}
      {[}{{{\color{delim}{[}}}}{1}
      {]}{{{\color{delim}{]}}}}{1},
}

\usepackage{xcolor}
\usepackage{listings}
\usepackage{enumerate}
\usepackage{siunitx}
\usepackage{booktabs}
\usepackage{hyperref}
\hypersetup{
     colorlinks   = true,
     citecolor    = blue
}
\newcommand{\ket}[1]{\left| #1 \right>} % for Dirac bras
\newcommand{\fix}[1]{{\color{red}#1}}
\definecolor{red}{rgb}{1,0.,0}

\usepackage[compatibility=false]{caption}
\usepackage{tcolorbox}

\begin{document}

\title{Extending C\texttt{++} for Heterogeneous Quantum-Classical Computing}

\thanks{This manuscript has been authored by UT-Battelle, LLC under Contract No. DE-AC05-00OR22725 with the U.S. Department of Energy. The United States Government retains and the publisher, by accepting the article for publication, acknowledges that the United States Government retains a non-exclusive, paid-up, irrevocable, world-wide license to publish or reproduce the published form of this manuscript, or allow others to do so, for United States Government purposes. The Department of Energy will provide public access to these results of federally sponsored research in accordance with the DOE Public Access Plan. (http://energy.gov/downloads/doe-public-access-plan).}

\begin{abstract}
We present \texttt{qcor} - a language extension to C\texttt{++} and compiler implementation that enables heterogeneous quantum-classical programming, compilation, and execution in a single-source context. Our work provides a first-of-its-kind C\texttt{++} compiler enabling high-level quantum kernel (function) expression in a quantum-language agnostic manner, as well as a hardware-agnostic, retargetable compiler workflow targeting a number of physical and virtual quantum computing backends. \texttt{qcor} leverages novel Clang plugin interfaces and builds upon the XACC system-level quantum programming framework to provide a state-of-the-art integration mechanism for quantum-classical compilation that leverages the best from the community at-large. \texttt{qcor} translates quantum kernels ultimately to the XACC intermediate representation, and provides user-extensible hooks for quantum compilation routines like circuit optimization, analysis, and placement. This work details the overall architecture and compiler workflow for \texttt{qcor}, and provides a number of illuminating programming examples demonstrating its utility for near-term variational tasks, quantum algorithm expression, and feed-forward error correction schemes.
\end{abstract}

\author{Thien Nguyen}
\affiliation{Quantum Computing Institute,\ Oak\ Ridge\ National\ Laboratory,\
  Oak\ Ridge,\ TN,\ 37831,\ USA}
\affiliation{Computer Science and Mathematics Division,\ Oak\ Ridge\ National\ Laboratory,\ Oak\ Ridge,\ TN,\ 37831,\ USA}

\author{Anthony Santana}
\affiliation{Quantum Computing Institute,\ Oak\ Ridge\ National\ Laboratory,\
  Oak\ Ridge,\ TN,\ 37831,\ USA}
\affiliation{Computer Science and Mathematics Division,\ Oak\ Ridge\ National\ Laboratory,\ Oak\ Ridge,\ TN,\ 37831,\ USA}

\author{Tyler Kharazi}
\affiliation{Quantum Computing Institute,\ Oak\ Ridge\ National\ Laboratory,\
  Oak\ Ridge,\ TN,\ 37831,\ USA}
\affiliation{Computational Sciences and Engineering Division,\ Oak\ Ridge\ National\ Laboratory,\ Oak\ Ridge,\ TN,\ 37831,\ USA}

\author{Daniel Claudino}
\affiliation{Quantum Computing Institute,\ Oak\ Ridge\ National\ Laboratory,\
  Oak\ Ridge,\ TN,\ 37831,\ USA}
\affiliation{Computer Science and Mathematics Division,\ Oak\ Ridge\ National\ Laboratory,\ Oak\ Ridge,\ TN,\ 37831,\ USA}

\author{Hal Finkel}
\affiliation{Leadership Computing Facility,\ Argonne\ National\ Laboratory,\ Lemont\, IL,\ 60439,\ USA}

\author{Alexander J.\ McCaskey}
\email{mccaskeyaj@ornl.gov}
\affiliation{Quantum Computing Institute,\ Oak\ Ridge\ National\ Laboratory,\
  Oak\ Ridge,\ TN,\ 37831,\ USA}
\affiliation{Computer Science and Mathematics Division,\ Oak\ Ridge\ National\ Laboratory,\ Oak\ Ridge,\ TN,\ 37831,\ USA}

\maketitle

%Travis feedback - I think the layout and narrative are good. The design documentation and code snippets are nice too. The last section on VQE and Adapt could benefit from presenting results of use. Even small demo examples would benefit the presentation. In addition, you need much more background on other languages and frameworks – the third paragraph in the Intro starts to do this, but it needs to be explained in more detail.
 
%%%%%%%%%%%%%%%%%%%%%%%%%
\section{Introduction}
The recent availability of programmable quantum computers over the cloud has enabled a number of small-scale experimental demonstrations of algorithmic execution for pertinent scientific computing tasks \cite{Dumitrescu2018,McCaskey2019,Kandala2017,PhysRevA.98.032331,hamilton}. These demonstrations point toward a future computing landscape whereby classical and quantum computing resources may be used in a hybrid, heterogeneous manner to continue to progress the state-of-the-art with regards to simulation capability and scale. A future post-exascale, heterogeneous computing architecture enhanced with quantum accelerators or co-processors in a tightly integrated manner could enable large-scale simulation capabilities for a number of scientific fields such as chemistry, nuclear and high-energy physics, and machine learning. However, the novelty and utility of heterogeneous quantum-classical compute models will only be effective if there is an enabling software infrastructure that promotes efficiency, programmability, and extensibility. There is therefore a strong need to put forward novel software frameworks, programming languages, compilers, and tools that will enable tight integration of existing HPC resources and applications with future quantum computing hardware. 

%he recent availability of programmable quantum computers over the cloud has enabled a number of small-scale experimental demonstrations of algorithmic execution for pertinent scientific computing tasks \cite{Dumitrescu2018,McCaskey2019,Kandala2017,PhysRevA.98.032331,hamilton}. These demonstrations point toward a future computing landscape whereby classical and quantum compute resources may be used in a hybrid, heterogeneous manner to continue to progress the state-of-the-art with regards to simulation capability and scale. A future post-exascale, heterogeneous computing architecture composed of classical and quantum accelerators or co-processors could enable large-scale simulation capabilities for a number of scientific fields such as chemistry, nuclear and high-energy physics, and machine learning. As quantum hardware architectures progress and become more noise-resilient, it is expected that these novel computational devices will serve as an accelerator for existing classical high-performance computing (HPC) architectures. There is therefore a strong need to put forward novel software frameworks, programming languages, compilers, and tools that will enable tight integration of existing HPC resources and applications with future quantum computing hardware. 

C\texttt{++} has proven itself as a leading language within the high-performance scientific computing community for its portability, scalability and performance, multi-paradigm capabilities (generic, object-oriented, imperative), integration with other languages, and community support. It has been leveraged to enable a number of programming models for classical accelerated computing \cite{cuda, sycl, raja, kokkos}. We anticipate that this trend will continue, and one will require models, compilers, and tools that promote node-level quantum acceleration via extensions or libraries for C\texttt{++}. Moreover, as tighter integration models become possible, quantum-classical programs that require a feed-forward capability (e.g. quantum error correction schemes) will require performant languages with low overhead. 

%As of this writing, a number of enabling software frameworks for quantum computing have been put forward, with most coming from quantum hardware vendors and provided as high-level Pythonic frameworks \cite{Qiskit, cirq-github, pyquil, projectQ, strawberryfields}. This works well for current remote execution models, but will prove problematic for the overall performance of quantum-classical interactions as quantum hardware progresses and tighter integration is enabled. Furthermore, quantum language and compiler approaches have been developed and deployed, but currently lack integration with standard HPC languages, software stacks, and workflows \cite{qsharp}, are not designed for high-level algorithmic expression \cite{jaqal, quil, openqasm}, or lack tight integration with physical backends \cite{scaffold}. On this latter point, a number of language compilers output optimized quantum assembly strings and execution of that code is left as a manual process for the programmer (mapping it to an appropriate data structure in the desired vendor-supplied Pythonic framework).
As of this writing, a number of approaches for programming quantum computers have been put forward, and one can classify most of these as either low-level intermediate or assembly languages, circuit construction frameworks, or high-level languages and compilers. Low-level intermediate languages like OpenQasm \cite{openqasm}, Quil \cite{quil}, and Jaqal \cite{jaqal} have been proposed that enable circuit definition at the gate or pulse level and most provide some form of hierarchical function (subroutine) definition, composition, and control-flow. Each of these provide its own set of benefits and drawbacks, most target a single hardware backend, and all are at a low-level of abstraction and are primarily meant to be generated by higher-level compilers and frameworks. Moving up the stack, there have been a number of Pythonic circuit construction frameworks developed (Qiskit \cite{Qiskit}, PyQuil \cite{pyquil}, Cirq \cite{cirq-github}, JaqalPaq \cite{jaqalpaq}, ProjectQ \cite{projectQ}) that make it easier for users to generate hardware-specific intermediate language representations for ultimate execution on remotely hosted backends. As hardware progresses and tighter CPU-QPU integration is enabled, we anticipate that this remote Pythonic programming and execution model will not be sufficient for enabling a performant interplay between classical and quantum resources. At the highest level, a few approaches have enabled high-level stand-alone, as well as embedded, domain specific languages and associated compilers for quantum-classical programming. We specifically look to Q\# \cite{qsharp} and Scaffold \cite{scaffold} as prototypical examples that have seen adoption and success. These approaches enable high-level expressibility as well as quantum-classical control flow. Unfortunately, both of these currently lack in some form with regards to tight integration of HPC resources with quantum co-processors. Q\# leverages the Microsoft .NET infrastructure and integrates with the C\# language, both of which are not easily adopted or accessed by existing HPC applications and resources. Scaffold extends C, a popular HPC language, but lacks direct integration with QPU resources, relying on manual processes for mapping compiler assembly output to appropriate Pythonic circuit-construction frameworks.

Here we describe a mechanism that seeks to fill this void in the quantum scientific computing software stack. Specifically, we detail the \texttt{qcor} compiler, which enables a language extension to C\texttt{++} through high-level Clang plugin implementations promoting quantum function expression alongside standard classical code. Our approach targets both near-term, remotely hosted quantum computing models as well as future fault-tolerant, tightly integrated quantum-classical architectures with feed-forward capabilities. We enable quantum code expression in a language agnostic manner as well as the ability to compile to most available quantum computing backends (including simulators). Furthermore, we provide a compiler runtime library that exposes a robust API for leveraging quantum kernels (functions) as standard functors or callables, to be leveraged as input to algorithmic implementations as needed. Ultimately, the \texttt{qcor} compiler paves the way for direct integration with existing applications, toolchains, and techniques common to scientific HPC, and is the first platform that allows programming hybrid quantum-classical algorithms in a single-source C\texttt{++}, general, and deployable manner.

This paper is outlined as follows: first we provide a quick discussion of a typical \texttt{qcor} program in an effort to guide the reader through the rest of the architectural details. We then provide the necessary background information required for a proper discussion of the \texttt{qcor} implementation (the specification, XACC, and Clang). Next, we provide the architectural details of the \texttt{qcor} runtime library and compiler implementation and workflow. The runtime library provides crucial utilities underpinning the language extension and compiler, as well as data structures and API calls for typical quantum algorithmic expression and execution. We detail the novel extensions to Clang we have developed for mapping general quantum kernel domain specific languages to valid C\texttt{++} API calls. We end with a robust demonstration of \texttt{qcor}, and demonstrate the programming of prototypical use cases, as well as its capability as an optimizing, retargetable quantum compiler.

\section{Anatomy of a QCOR Program}
\begin{figure}[t!] 
  \lstset {language=C++}
  \begin{lstlisting}
// No includes needed, we are using the 
// language extension

// Quantum Kernels are just C++ functions
// annotated with __qpu__. Can take any arguments
// must provide a qreg to run on.
__qpu__ void bell(qreg q) {
  // Kernels can be expressed in any available 
  // quantum language, here XACC XASM.
  // The language extension allows quantum 
  // instruction expression as part of the language
  H(q[0]);
  CX(q[0], q[1]);
  
  // but we also get control flow for free
  for (int i = 0; i < 2; i++) {
    Measure(q[i]);
  }
}

// Just standard C++ 
int main() {
  // Language extension gives us the 
  // qalloc() quantum buffer allocator. 
  // q is a qreg, a primitive type provided 
  // by the language extension
  auto q = qalloc(2);
  
  // Execute the quantum kernel by just calling it
  bell(q);
  
  // Results are available on the allocated qreg
  q.print();
}
// Run on remote IBM Paris backend with 
// qcor -qpu ibm:ibmq_paris -shots 1024 \
//        bell.cpp -o bell.x
// ./bell.x
\end{lstlisting}
\caption{The simplest \texttt{qcor} program, expressing a quantum kernel that executes the standard Bell state.}
\label{fig:qcor_simple}
\end{figure}
Figure \ref{fig:qcor_simple} demonstrates a simple \texttt{qcor}-enabled C++ program - the programming and execution of the Bell state. This straightforward case demonstrates the single-source programming model \texttt{qcor} provides, without going into all the complexity in the rest of the \texttt{qcor} / XACC framework for common algorithmic tasks. We will go into the full details of the \texttt{qcor} implementation in the following sections, but here we show the model and the philosophy put forward by the language extension.

Critically, the \texttt{qcor} compiler enables a C\texttt{++} language extension that enables the use of a primitive \texttt{qreg} type, quantum kernel definition, primitive quantum instruction programming, and quantum-classical control flow. In the code snippet, one notices there are no header files included, everything in the source code is provided by the language extension. Programmers begin by defining a quantum kernel, which is just a standard C\texttt{++} function annotated with the \texttt{\_\_qpu\_\_} attribute. Kernels can take arbitrary function arguments, but must take at least one reference to an allocated qubit register (\texttt{qreg}). The function body itself is language-agnostic, i.e., programmers can use any quantum programming language (for which there is an appropriate \texttt{TokenCollector} implementation, see Section \ref{sec:qcor_syntax}). The current version of \texttt{qcor} enables one to program in the XASM \cite{xasm2000}, IBM OpenQasm \cite{openqasm}, Quil \cite{quil}, and  custom unitary matrix decomposition languages. Notice that low-level quantum instruction invocation is allowed as part of the language extension itself, and that we are free to use existing C\texttt{++} control flow statements like the \texttt{for} loop used to apply measurement instructions. Once the kernel is defined, one simply allocates a register of qubits of a desired size (similar to the C \texttt{malloc} call but for qubits, \texttt{qalloc}). To execute the kernel on the targeted quantum co-processor, one just invokes the quantum kernel function, providing the correct arguments (here the qubit register). Execution results (bit strings and counts) are persisted to the \texttt{qreg} instance and are available for use in the rest of the program. 

To compile and run this program, one uses the \texttt{qcor} compiler, indicating the quantum backend being compiled to and any other pertinent execution information (like shots). The \texttt{qcor} compiler provides all of the same compiler command line arguments as Clang and GCC, i.e., one can build up complex source codes that require extra header and library search paths, specific libraries to link, and other compiler and link flags. After compilation, the programmer is left with a binary executable or object file. 

Figure \ref{fig:qcor_simple} is a simple example of programming with \texttt{qcor}. There is of course much more that one could do, including kernel composition (kernels that call other kernels), auto-generated adjoint and control versions of the defined quantum kernel, kernel construction with complex control flow, kernel definition at the unitary matrix level via extensible circuit synthesis algorithms, and the use of \texttt{qcor} provided data structures for the expression of complex hybrid quantum-classical algorithms. The rest of this work will describe these key abilities in the following sections. 

\section{Background}
\begin{figure}[b!]
\centering  
\includegraphics[width=0.5\textwidth]{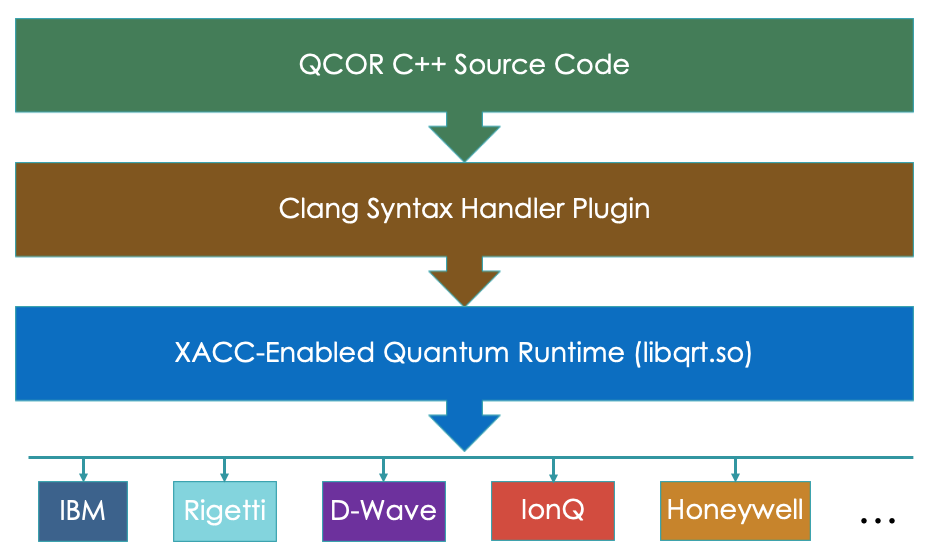}
\caption{\texttt{qcor} provides a single-source C\texttt{++} programming model through plugin extensions to Clang and an XACC-enabled quantum runtime library implementation, enabling execution on a number of popular quantum backends.}
\label{fig:qcor_layers}
\end{figure}
\texttt{qcor} implements the specification put forward in \cite{mintz2019qcor} by building upon the XACC quantum programming framework. Moreover, quantum kernel compilation is accomplished via extension of core Clang plugin interfaces. Here we describe pertinent details about XACC, the QCOR specification, and Clang in order to provide a foundation to describe the \texttt{qcor} compiler implementation. Figure \ref{fig:qcor_layers} gives a high-level view of the overall relationship between QCOR, Clang, and XACC. QCOR kernel expressions are mapped to appropriate XACC types via domain-specific language preprocessing provided by novel plugins to the Clang infrastructure. The incorporation of XACC implies a retargetable compiler workflow, with backends provided by the main quantum computing hardware vendors.

\subsection{XACC}
The XACC quantum programming framework is a system-level, C\texttt{++} infrastructure enabling language and hardware agnostic quantum programming, compilation, and execution \cite{mccaskey2020xacc}. XACC adopts a dual-source programming model, whereby quantum kernels are defined as separate source strings and compiled to a core, polymorphic intermediate representation (IR) via an appropriate API library call. XACC builds upon the CppMicroServices framework \cite{CppMicroServices} to provide a native implementation of the Open Services Gateway Initiative (OSGi) \cite{osgi}, and promote a service oriented architecture that provides extensibility at all points of the quantum-classical programming workflow. We leave a detailed overview of XACC to the seminal paper \cite{mccaskey2020xacc}, but here we highlight a few core service interfaces that are pertinent for our discussion of \texttt{qcor}.

XACC employs a layered architecture that decomposes the framework into extensible frontend, middle-end, and backend layers. The frontend exposes a service interface, the \texttt{Compiler}, that maps kernel source strings to instances of the IR, in a language-specific manner. The middle-end exposes extension points defining the quantum intermediate representation, which is a polymorphic object model for representing compiled quantum kernels. It is composed of \texttt{Instruction} and \texttt{CompositeInstruction} service interfaces which, for gate model computing, are specialized for concrete quantum gates and composites of those gates, respectively. The middle-end also exposes an \texttt{IRTransformation} service interface that enables the general transformation of \texttt{CompositeInstructions}, important for quantum compilation tasks such as general circuit optimization, low-level synthesis, analysis, and circuit placement. Finally, the backend layer exposes an extensible interface for injecting physical and virtual quantum computing backends - the \texttt{Accelerator}. XACC puts forward another critical concept for modeling an allocation of quantum memory (a register of qubits) called the \texttt{AcceleratorBuffer}. This data structure spans the three architectural layers and is instantiated by programmers and passed to backend Accelerators for execution - we say \texttt{Accelerators} execute \texttt{CompositeInstructions} on a given \texttt{AcceleratorBuffer}. The results of execution are persisted to the buffer and immediately available to the programmer that instantiated, and still has reference to, that buffer.

These core concepts - kernel \texttt{Compilers}, \texttt{Instructions} and \texttt{CompositeInstructions}, \texttt{IRTransformations}, \texttt{Accelerators}, and \texttt{AcceleratorBuffers} - make up the key elements that will be leveraged in our single-source C\texttt{++} programming model and language extension implementation. High-level quantum kernels in \texttt{qcor} will have a corresponding \texttt{CompositeInstruction} instance that will be generated by variants of the \texttt{Compiler} service. Quantum compilation optimization and placement routines will be injected as implementations of the \texttt{IRTransformation}. The retargetability of the compiler will be due to the interchangeable characteristic of backend \texttt{Accelerators}. The language extension representation of a register of qubits, or \emph{qreg}, will be represented under the hood as an \texttt{AcceleratorBuffer}. 

\subsection{QCOR Specification}
The language extension specification put forward in \cite{mintz2019qcor} defines a single-source programming model for heterogeneous quantum-classical quantum computing that leverages a shared memory model and an asynchronous task-based execution model. Moreover, it puts forward a data-model that provides a set of abstractions for describing general hybrid quantum-classical variational algorithms for near-term quantum computation. The \texttt{qcor} compiler implementation, in tandem with XACC, implements this specification for the case of extending the C\texttt{++} programming language. The data model specification puts forward the \texttt{Operator}, \texttt{Optimizer}, and \texttt{ObjectiveFunction} abstractions for composing hybrid variational algorithms, and the \texttt{taskInitiate()} call for asynchronous execution. \texttt{Operators} represent quantum mechanical operators or compositions of operators that can \texttt{observe} unmeasured quantum kernels (if the \texttt{Operator} is Hermitian), returning a list of measured quantum kernels. An example of this would be an \texttt{Operator} sub-type representing Pauli operators or sums of Pauli tensor products. The \texttt{Optimizer} concept represents a multi-variate function optimization strategy (COBYLA \cite{cobyla}, L-BFGS \cite{lbfgsb}, Adam \cite{adam}, etc.). We have provided implementations of the \texttt{Operator} and \texttt{Optimizer} as part of the latest release of XACC \cite{mccaskey2020xacc}. The \texttt{ObjectiveFunction} concept represents a multi-variate function that returns a scalar value, and evaluation of the function to produce that scalar requires quantum co-processor execution. An example of this would be the variational quantum eigensolver (VQE) workflow, where one has a parameterized circuit and would like to execute the circuit and evaluate the expectation value of some \texttt{Operator}. Finally, the specification stipulates a \texttt{taskInitiate()} API call and associated overloads that will execute a hybrid quantum-classical task asynchronously, enabling the host thread to continue classical processing in parallel. 

\subsection{Clang Plugins}
We base our \texttt{qcor} compiler implementation upon the Clang compiler frontend infrastructure due to its excellent support and utility in academia and industry, its overall extensibility and modularity, and its ability to enable the injection of custom plugin implementations for various aspects of the compiler frontend workflow.

Clang is the C\texttt{++} frontend for the LLVM compiler infrastructure~\cite{lattner2004llvm}, responsible for converting C\texttt{++} source code into LLVM's intermediate representation. Clang uses LLVM to compile C\texttt{++} source code to executable objects, and in addition, can perform tasks such as static analysis and source rewriting. At a high-level, the Clang infrastructure puts forward a robust object model for lexing, parsing, preprocessing, abstract syntax tree (AST) generation, and LLVM-IR code generation. Clang supports several plugin interfaces that can be used, in arbitrary combination, to enhance Clang's ability to process C\texttt{++} source code. Existing plugin interfaces are the \texttt{ASTConsumer}, allowing a plugin to monitor the creation of AST nodes, and the \texttt{PragmaHandler}, allowing a plugin to process custom pragma directives. Plugins, in general, have access to Clang's AST data structures and the state describing how an individual C\texttt{++}source file is being compiled.

\begin{figure}[t!] 
  \lstset {language=C++}
  \begin{lstlisting}
[[clang::syntax(sh_name)]] void foo() {
  ... Embedded DSL here
  ... SyntaxHandler with name sh_name will 
  ... translate this to standard C++ code
}
---------------------------------------------------
using namespace clang;
using namespace llvm;
class MySyntaxHandler : public SyntaxHandler {
public:
  MySyntaxHandler() : SyntaxHandler("sh_name") {}
  void GetReplacement(Preprocessor& PP, 
                   Declarator& D, 
                   CachedTokens& Toks, 
                   raw_string_ostream& OS) override 
  { 
     ... analyze Toks, write new code to OS
  }
  void AddToPredefines(raw_string_ostream& OS) {
    ... add any #includes here
  }
};
\end{lstlisting}
\caption{Demonstration of how the Clang SyntaxHandler works. Programmers annotate a function indicating the SyntaxHandler to be used in parsing and transforming the function body Tokens.}
\label{fig:clang_sh}
\end{figure}

An early design goal of this work that separates it from others in the field is to ensure that all Clang extensions for enabling \texttt{qcor} functionality and features are contributed as separate plugin implementations. We explicitly avoid making core, permanent modifications to the core of Clang or LLVM. Doing so would force us to maintain a separate fork of these huge code-bases. We adopt the simpler route - extend key points of the preprocessing workflow with custom plugin implementations, and enable users to build \texttt{qcor} off existing Clang/LLVM binary installs.

To this effect, \texttt{qcor} makes use of a newly-proposed plugin interface: the syntax handler, \texttt{SyntaxHandler}~\cite{clangsh}. The syntax handler allows embedding of domain-specific languages into C\texttt{++} function definitions. Each syntax handler implementation (see Figure \ref{fig:clang_sh}) registers to handle a specific, named syntax tag. Functions with the C\texttt{++} attribute \texttt{[[clang::syntax(\textit{tag})]]} are processed by Clang's parser is a special way. First, the function body is extracted by collecting all tokens prior to the closing '\}' using balanced-delimiter matching. Thus, while the text in the body of the function does not need to be valid C\texttt{++} code, it is subject to C\texttt{++} preprocessing and cannot contain unbalanced '\{' and '\}' characters. The token stream is then provided to the syntax-handler plugin along with information about the already-parsed function declarator. The declarator contains information about the function's name and arguments. The plugin provides, in return, a replacement text stream for the function. This text stream is then subjected to tokenization, much in the same way as an included source file might be handled, and parsing continues using the replacement text instead of the original function body. As described in Section \ref{sec:qcor_syntax}), we leverage this plugin interface to translate our quantum kernel expressions to valid C\texttt{++} API calls. 

\section{QCOR}
Now we turn to the internal architecture that enables the functionality put forward by the QCOR specification. Ultimately, our \texttt{qcor} compiler implementation is composed of a runtime library as well as a Clang \texttt{SyntaxHandler} implementation enabling compilation of quantum kernels to valid C\texttt{++} API calls (specifically, calls to the runtime library, and ultimately XACC). The runtime library puts forward a number of key abstractions that implement the original specification. Specifically, the runtime library provides a \texttt{QuantumKernel} class abstraction, implementations of \texttt{ObjectiveFunction}, \texttt{Operator}, and \texttt{Optimizer}, a novel quantum runtime library API, and a task-based asynchronous execution API. The compiler provides a Clang \texttt{SyntaxHandler} that ensures quantum kernel domain specific languages (invalid code with respect to other compilers) are mapped to appropriate and valid sub-types of the \texttt{QuantumKernel} abstraction, as well as other utility functions. This mechanism ensures the quantum language-agnostic characteristic of our specification and implementation. The compiler module of \texttt{qcor} currently enables programming in XASM, OpenQasm, and Quil, as well as a custom language for expression unitary matrices to be decomposed into quantum assembly.  

\subsection{Runtime}
\subsubsection{Quantum Kernel}
The QCOR specification stipulates that the quantum kernel must be some functor-like object with a function body composed of quantum code provided in some domain specific language, and execution of the functor affects execution of that quantum expression on the quantum co-processor. Beyond that, the specification currently allows language extension implementors to freely describe the kernel object model in a way that best suits the language being extended. For our \texttt{qcor} compiler implementation, we specify quantum kernels as C\texttt{++} functions that are annotated with a \texttt{\_\_qpu\_\_} attribute, return void, and can take any function arguments, with at least one \texttt{qreg} argument. The function body can contain quantum code expressions written in any available quantum language. Here the word \emph{available} implies the compiler has an appropriate token analysis implementation for the quantum domain specific language. 

\begin{figure}[t!]
\centering  
\includegraphics[width=.45\textwidth]{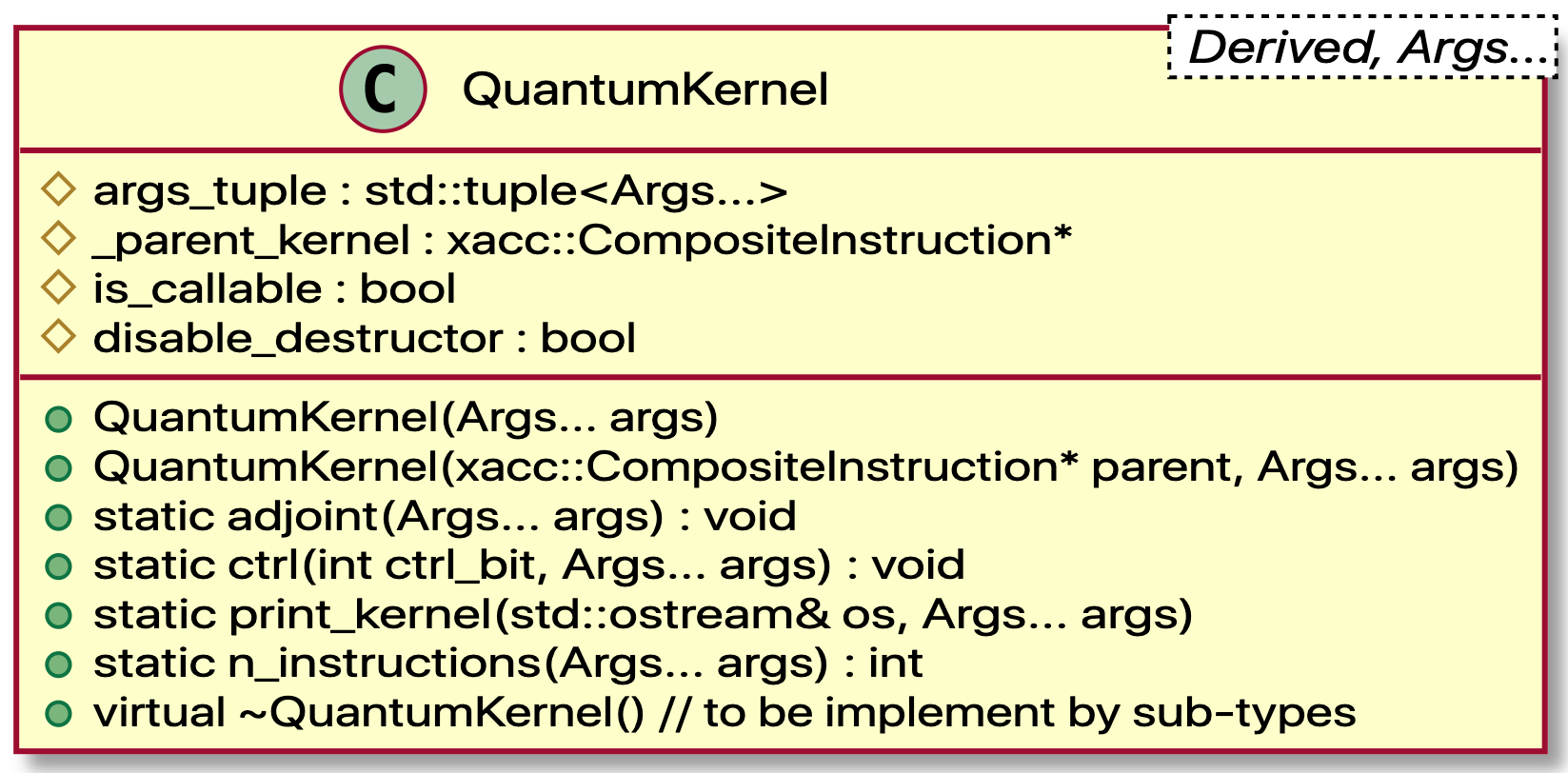}
\caption{The class diagram for the \texttt{QuantumKernel} template class. This class exposes constructors for entry-point kernels, kernel composition, and static methods for the generation of related circuits.}
\label{fig:qkernel_uml}
\end{figure}

\begin{figure}[t!] 
  \lstset {language=C++}
  \begin{lstlisting}
__qpu__ void ansatz(qreg q, double x) {
  X(q[0]);
  Ry(q[1], x);
  CX(q[1], q[0]);
}
... representation as QuantumKernel sub-type ...
class ansatz : 
        public QuantumKernel<ansatz, 
                            qreg, double> {
protected:
  void operator()(qreg q, double x) {
    // fill _parent_kernel
    // add x, ry, cx using QuantumRuntime
  }
public:
  ~ansatz_z0z1() {
    auto [q,x] = args_tuple;
    operator()(q,x);
    // submit _parent_kernel via QuantumRuntime
  }
}
... instantiating a temp instance
... looks like evaluation 
ansatz(q, 2.2);
// can also use auto-generated static methods
ansatz::adjoint(q,2.2);
ansatz::ctrl(1, q, 2.2);
\end{lstlisting}
\caption{Code snippet demonstrating how a quantum kernel gets represented as a \texttt{QuantumKernel} sub-type.}
\label{fig:qkernel_snippet}
\end{figure}
In order to represent this kernel concept as part of the runtime library, \texttt{qcor} exposes a \texttt{QuantumKernel} class that follows the familiar curiously-recurring template pattern (CRTP) \cite{crtp} and is intended to serve as a super-type for concrete kernel implementations. It takes the type of the subclass as its first template argument (\texttt{Derived} in Figure \ref{fig:qkernel_uml}), followed by a variadic template parameter pack describing the quantum kernel function argument types (\texttt{Args...} in Figure \ref{fig:qkernel_uml}). The class keeps reference to a \texttt{std::tuple} on the variadic types and stores concrete function argument instances in the tuple upon construction (the first constructor in Figure \ref{fig:qkernel_uml}). Crucially, the class also keeps reference to an \texttt{xacc::CompositeInstruction} pointer (the \texttt{\_parent\_kernel} member) - an internal representation of this quantum kernel as an XACC IR instance. This is used for ultimate submission to the quantum co-processor (an instance of the XACC \texttt{Accelerator}). To promote quantum kernel composition (kernels that call other kernels), \texttt{QuantumKernel} exposes a second constructor that takes an upstream \texttt{xacc::CompositeInstruction} pointer. So an entry-point kernel (a quantum kernel called from a classical function) can work to fill its \texttt{\_parent\_kernel} instance, and then pass that to another kernel instance for it to use as its internal \texttt{\_parent\_kernel}. This pattern directly enables quantum kernel composition.

The \texttt{QuantumKernel} class is never intended for use on its own, but rather it is meant to be subclassed by concrete quantum kernel representations. The design strategy for sub-types is to inherit from \texttt{QuantumKernel}, passing the sub-type itself as the first template argument, followed by the kernel function argument types, then provide an implementation of the sub-type destructor that ultimately affects execution of the quantum code. Figure \ref{fig:qkernel_snippet} demonstrates this, where we have a parameterized quantum kernel, \texttt{ansatz}, that takes a \texttt{qreg} and \texttt{double} parameter. We subclass \texttt{QuantumKernel<ansatz, qreg, double>} and provide a means for execution at destruction. Specifically, the sub-type should fill the \texttt{\_parent\_kernel} \texttt{CompositeInstruction} and submit for execution. By doing this, one can see that instantiating a temporary instance of \texttt{ansatz} looks like quantum kernel function evaluation. 

By doing it this way, we allow ourselves the opportunity to provide extra functionality for quantum kernels that you could not get through a standard function alone. For example, defining the \texttt{QuantumKernel} class gives us an opportunity to define extra public class methods that enable pertinent analysis tasks, like printing the kernel to an output stream or viewing depth, number of gates, or other circuit-specific information. Moreover, this gives us the opportunity to automatically generate related circuits. Figure \ref{fig:qkernel_uml} shows two such static methods, \texttt{adjoint} and \texttt{ctrl}, which auto-generate the adjoint / reverse and controlled version of the given quantum kernel automatically. 

We do not expect the average \texttt{qcor} user to be concerned too much with the \texttt{QuantumKernel} class. It is primarily intended to serve as an internal representation of the quantum kernel that enables high-level programmability, as well as provide extra internal features for compiler and library developers. The primary goal of the \texttt{qcor} compiler is to map quantum kernel functions to appropriate definitions of \texttt{QuantumKernel} sub-types. 

\subsubsection{Quantum Runtime}
\begin{figure}[t!]
\centering  
\includegraphics[width=.5\textwidth]{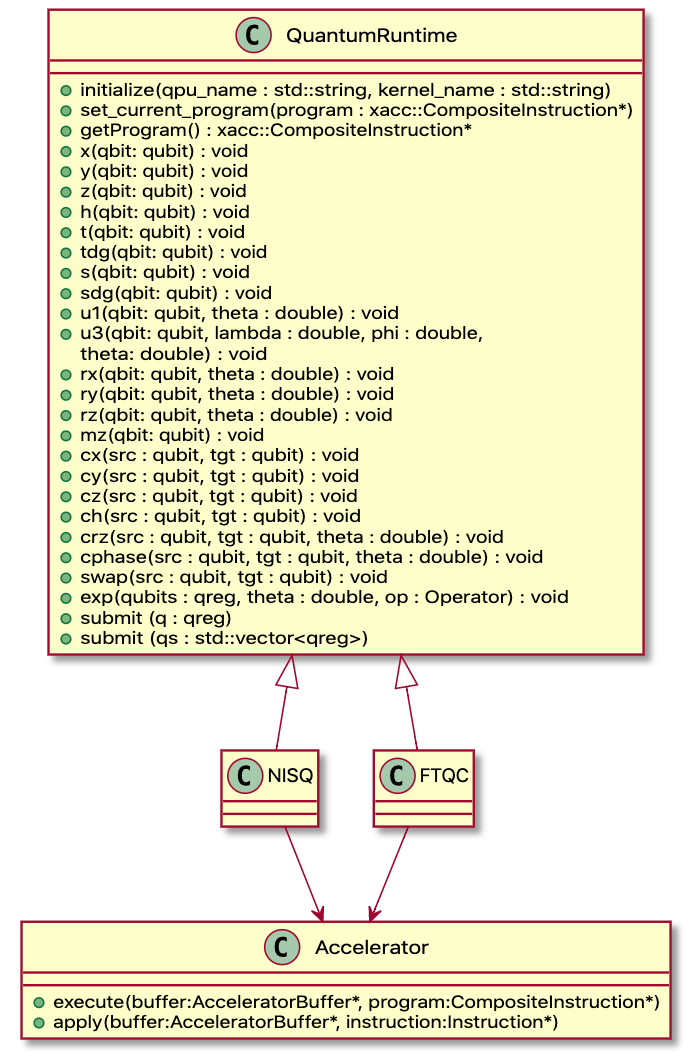}
\caption{The class diagram for the \texttt{QuantumRuntime} class. We provide implementations of this that enable both remotely hosted QPU execution, as well as future fault-tolerant models that stream instruction execution on a tightly integrated quantum backend.}
\label{fig:qrt_uml}
\end{figure}

The \texttt{qcor} \texttt{QuantumRuntime} exposes a class API for compiler and runtime developers to execute low-level quantum gate instructions on the specified quantum backend. This class represents a critical piece of the \texttt{qcor} runtime library architecture in that it provides an extensible hardware abstraction layer enabling typical quantum instruction execution. Moreover, it promotes the utility of different models of quantum-classical integration - remote, near-term models as well as tightly integrated feed-forward models. For near-term applications, the \texttt{QuantumRuntime} can be implemented to queue gate instructions as their corresponding API call is invoked. This execution paradigm keeps track of an internal representation of the low-level quantum circuit, and for each gate-level API invocation in a given quantum kernel execution context, the internal representation is built up, effectively queuing each instruction as it comes in. At the end of this construction or queuing period, the API exposes a \texttt{submit()} call that flushes the internal representation, sending the entirety of its contents to be executed on the compiled backend. This is demonstrated in Figure \ref{fig:qrt_uml} as the \texttt{NISQ} subtype, and is the default \texttt{QuantumRuntime} backend in \texttt{qcor}. 
Specifically, this default implementation of the \texttt{QuantumRuntime} API keeps track of a \texttt{xacc::CompositeInstruction} member that it populates upon each invocation of a quantum gate function call. Ultimately, the \texttt{QuantumRuntime} API represents a public interface for constructing XACC IR instances programmatically. The \texttt{QuantumRuntime} exposes methods for all common single qubit (Hadamard, T, S, Tdg, Sdg, Rx, Ry, Rz, U3, U1, X, Y, Z),  two qubit (CX, CY, CZ, CH, CPhase, CRz, Swap), and measurement gates, as well as more complicated circuit synthesis routines like a function for first order trotterization of a provided \texttt{qcor} \texttt{Operator} (\texttt{exp()} function call). The default \texttt{submit} call takes an \texttt{qreg} instance and configures the execution of the \texttt{xacc::CompositeInstruction} on the backed \texttt{xacc::Accelerator} specified at compile time. 

To support quantum hardware capable of fast feedback between the quantum processor and the classical processor, we also put forward a fault-tolerant quantum runtime (\texttt{FTQC} subtype as shown in Figure~\ref{fig:qrt_uml}). In this execution model, the runtime library will dispatch quantum instructions to the \texttt{Accelerator} backend immediately and reflect any measurement results to the classical code as return values of the \texttt{QuantumRuntime::mz()} function. This \texttt{FTQC} runtime enables flexible control flow of our quantum kernels such as that required for quantum error correction implementations, whereby syndrome decoding is performed in real-time by a classical computer to determine appropriate correction strategies.

Since our provided \texttt{QuantumRuntime} implementations default to XACC, \texttt{qcor} picks up support for a number of physical quantum computers via the XACC \texttt{Accelerator} extension point, which ultimately handles mapping the XACC IR to the appropriate native gate set. However, one further design goal of this interface is to enable developers to extend the \texttt{QuantumRuntime} with a more robust level of support for the designated backend. We anticipate that this interface may enable implementations for specific physical backends, or even for lower-level electronic control system APIs, that provide a more efficient IR-translation mechanism for native backend gate sets. 

\subsubsection{Operator, Optimizer, and Objective Function}
\begin{figure}[b!] 
  \lstset {language=C++}
  \begin{lstlisting}
// Create Operator from string
auto H = createOperator("pauli", 
                "2.2 X0 X1 + 3.3 Y0 Y1");
// Create Operator from X, Y, Z, API
auto H = 5.907 - 2.1433 * X(0) * X(1) -
           2.1433 * Y(0) * Y(1) + .21829 * Z(0) -
           6.125 * Z(1);
// Create from a, adag API
auto H = adag(1) * a(0) + adag(0) * a(1);
// Create from Operator Generators
auto H2_chem =
    createOperator("chemistry", 
     {{"basis", "sto-3g"}, {"geometry", H2_geom}});
    
// Create Optimizer based on NLOPT (COBYLA default)
auto optimizer = createOptimizer("nlopt");
// Create Adam from MLPACK
auto optimizer = createOptimizer("mlpack", 
                {{"mlpack-optimizer", "adam"}});
\end{lstlisting}
\caption{Demonstration of creating and using \texttt{qcor} Operators and Optimizers.}
\label{fig:obs_opt_code}
\end{figure}
The QCOR specification defines a few concepts that seek to enable efficient expression of common quantum algorithms, specifically those that are variational and target potential near-term quantum hardware. These types, the \texttt{Operator}, \texttt{ObjectiveFunction}, and \texttt{Optimizer}, provide the necessary abstractions at a familiar level to enable general variational tasks that leverage quantum co-processing. The \texttt{qcor} implementation seeks to enable these concepts in a manner that is modular and extensible, allowing future \texttt{qcor} developers to tailor these concepts to their specific workflow. 

First, the \texttt{Operator} concept represents a general quantum mechanical operator, or composition of operators. The \texttt{Operator} should expose appropriate algebra that enables programmers to build up complicated Hamiltonian models that can be leveraged for quantum simulation. Critically, \texttt{Operators} must expose some mechanism for the \emph{observation} of quantum states on the quantum co-processor. By this we mean, given some unmeasured quantum kernel, the \texttt{Operator} should return a list of measured kernels, dependent solely on its internal structure. The prototypical example of this would be the VQE algorithm, whereby you have an \texttt{Operator} that describes the Hamiltonian of interest consisting of a sum of Pauli tensor products, and one requires quantum kernel executions for each term followed by measurements in the basis of the term itself. \texttt{qcor} implements the \texttt{Operator} concept as a class to be sub-typed for specific quantum mechanical operator types, each encoding its own operator algebra. The class exposes an interface for algebra (appropriate operator overloads in C\texttt{++}), as well as common methods for operator analysis. Every \texttt{Operator} in \texttt{qcor} can be instantiated from string, from a site-map (qubit index to operator name), or from a mapping of options. 
\begin{figure*}[!p]
\centering  
\includegraphics[width=\textwidth]{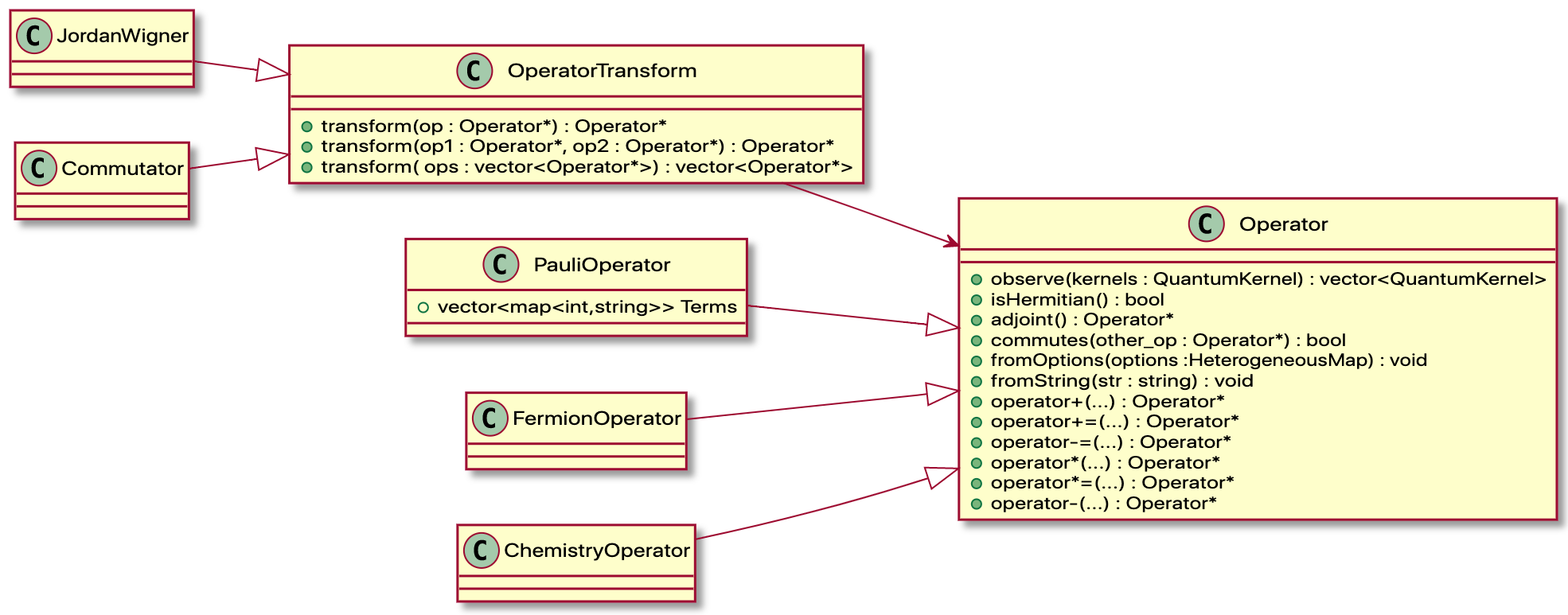}
\caption{The class diagram for the \texttt{Operator} class. \texttt{Operator} exposes an API for algebraic operations, which sub-types implement.}
\label{fig:qcor_operator_uml}
\end{figure*}
\begin{figure*}[p!]
\centering  
\includegraphics[width=\textwidth]{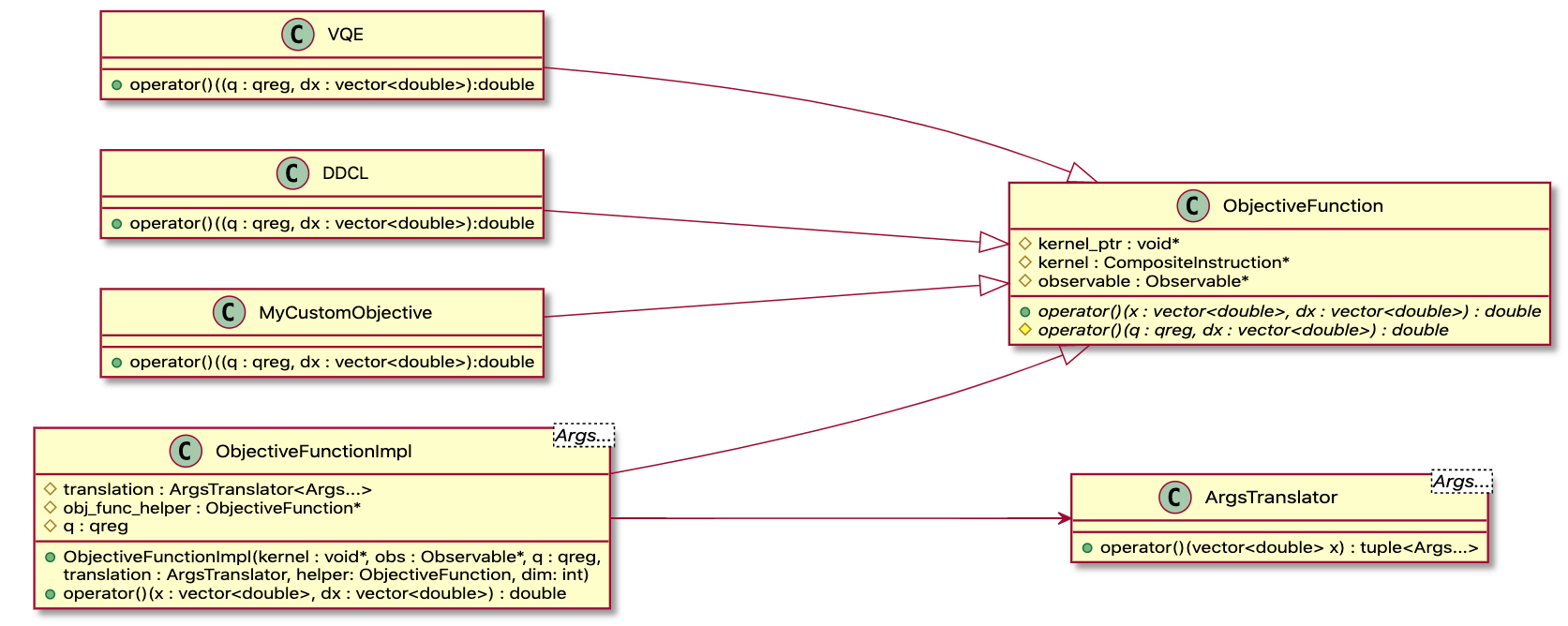}
\caption{The class diagram for the \texttt{ObjectiveFunction} template class. Sub-types provide custom \texttt{ObjectiveFunction} evaluation workflows.}
\label{fig:obj_func_uml}
\end{figure*}
\begin{figure*}[!p]
\centering  
\includegraphics[width=\textwidth]{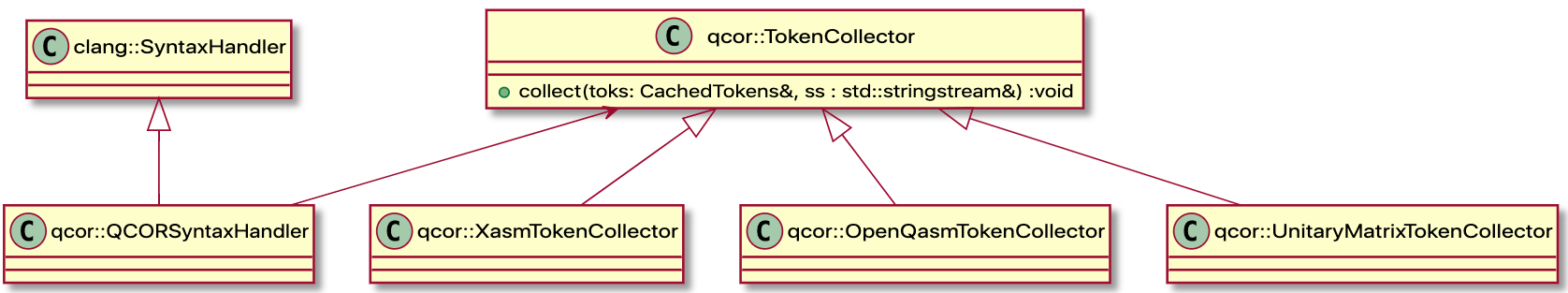}
\caption{The class diagram for the \texttt{QCORSyntaxHandler} class.}
\label{fig:qcor_sh_uml}
\end{figure*}
\clearpage
\texttt{qcor} provides sub-types for Pauli and Fermionic operators, as well as more complicated \texttt{Operators} that auto-generate themselves from this mapping of options (e.g., molecular geometry and basis set name to generate a molecular Hamiltonian, for example). \texttt{qcor} provides a creation API for \texttt{Operators} that enables efficient expression of quantum operators in a way that is familiar for programmers (see Figure \ref{fig:obs_opt_code}). The class architecture for the \texttt{Operator} and related types is shown in Figure \ref{fig:qcor_operator_uml}. Moreover, we further define and provide the \texttt{OperatorTransform} to serve as an extension point for general transformations on \texttt{Operators} (e.g., Jordan-Wigner for mapping Fermionic \texttt{Operators} to Pauli ones). 

The \texttt{ObjectiveFunction} concept in \texttt{qcor} represents a multi-variate function that returns a scalar value ($y = F(\textbf{x})$), and evaluation of the function requires quantum co-processor execution (e.g., the VQE workflow, where one has a parameterized circuit and would like to execute the circuit and evaluate the expectation value of some \texttt{Operator}). Ultimately, the \texttt{ObjectiveFunction} generalizes the notion of pre-processing, circuit evaluation, and post-processing in order to produce some scalar value given a vector of input scalar parameters. 
This concept has proven ubiquitous throughout near-term variational quantum-classical algorithm development and utility. In order to affect that workflow, \texttt{ObjectiveFunctions} requires initialization with both the quantum kernel of interest (passed as a functor or function pointer) and the \texttt{Operator} dictating measurements on the kernel. 
\begin{figure}[t!] 
  \lstset {language=C++}
  \begin{lstlisting}
__qpu__ void foo(qreg, double x) {
 .... quantum circuit using x parameter
}
...
auto H = createOperator("pauli", "X0 + Y1");
int n_params = 1;

// Create Objective to 
// evaluate <foo(x) | H | foo(x)>
auto objective = 
    createObjectiveFunction(foo, H, n_params);

// Evaluate at a concrete vector of parameters.
auto exp_val_H = (*objective)({1.345});

// Perform parameter sweep
for (auto x : linspace(-constants::pi, 
                        constants::pi, 20)) {
  std::cout << "Value at " << x << " is " << 
            (*objective)({x}) << "\n";
}
\end{lstlisting}
\caption{Demonstration of creating an \texttt{ObjectiveFunction} and using it for evaluation. Here we demonstrate the default VQE objective, returning the expected value of the provided \texttt{Operator}.}
\label{fig:createobj}
\end{figure}
The class architecture for the \texttt{ObjectiveFunction} is shown in Figure \ref{fig:obj_func_uml}, which we decompose into user-level \texttt{ObjectiveFunction} and internal \texttt{ObjectiveFunctionImpl} classes. The latter class is a variadic template on the quantum kernel argument types that keeps reference to an internal \emph{helper} \texttt{ObjectiveFunction} and implements the \texttt{operator()(std::vector<double>)} method to map the incoming parameter vector to appropriate quantum kernel function arguments. It then invokes the protected \texttt{operator()(qreg, std::vector<double>\&)} method of its \texttt{ObjectiveFunction} helper reference, passing the internal \texttt{qreg} instance and the reference to a vector for gradients, and returns the result of that call. 

Conceptually, programmers request an \texttt{ObjectiveFunction} (see Figure \ref{fig:createobj}) of a given name (\texttt{vqe} for example) and an \texttt{ObjectiveFunctionImpl} is constructed internally, templated on the quantum kernel arguments, and given reference to the correpsonding \texttt{ObjectiveFunction} instance as its \texttt{obj\_func\_helper}. The \texttt{ObjectiveFunctionImpl} is solely responsible for evaluation of the quantum kernel, with pre- and post-processing left as a job for the internal helper \texttt{ObjectiveFunction}. The \texttt{ObjectiveFunctionImpl} instance is returned to programmers as an \texttt{ObjectiveFunction} pointer, removing the need for users to know any information about the template types or internal implementation.

\begin{figure}[t!] 
  \lstset {language=C++}
  \begin{lstlisting}
// assume a kernel like this
__qpu__ void foo(qreg, std::vector<double> gamma, 
                    std::vector<double> beta) {
 .... quantum circuit using gamma, beta params
}
...
const int mid_point = 4;
auto args_translator = 
    ArgsTranslator<std::vector<double>, 
                        std::vector<double>>(
    [&](const std::vector<double> x) {
    // split x into gamma and beta sets
    std::vector<double> gamma(x.begin(), 
                    x.begin() + mid_point),
        beta(x.begin() + mid_point, x.end());
    return std::make_tuple(q, gamma, beta);
});
\end{lstlisting}
\caption{Demonstration of providing a custom \texttt{ArgsTranslator} that provides a mapping between the \texttt{ObjectiveFunction}'s requisite \texttt{std::vector<double> x} and complex kernel argument structures.}
\label{fig:tfunc_code}
\end{figure}
It should be noted that in advanced use cases, the quantum kernel argument signature may in general be much more complex than the argument signature of the \texttt{ObjectiveFunction} functor. In this case, we need a mechanism for mapping \texttt{std::vector<double> x} parameters to the argument structure of the provided quantum kernel. To achieve this, \texttt{qcor} defines the \texttt{ArgsTranslator<Args...>} variadic class. This concept is templated on the argument types of the quantum kernel, and takes at construction a lambda or functor of signature \texttt{std::tuple<Args...>(const std::vector<double>)}. The goal of this lambda is to map the incoming parameter vector to the kernel function arguments, taking advantage of any lambda capture variables required. A concrete example of this would be in the definition of a quantum kernel that takes two separate parameter vectors which, if concatenated together, would form the single parameter vector required for \texttt{ObjectiveFunction}. In this case, one would define a \texttt{ArgsTranslator} like in the code snippet provided in Figure \ref{fig:tfunc_code}. 

\texttt{qcor} provides a creation API for \texttt{ObjectiveFunctions}, \texttt{createObjectiveFunction()} (see Figure \ref{fig:createobj}), with a few overloads: (1) take as input a quantum kernel functor and an \texttt{Operator}, which defaults to evaluating the expectation value of the \texttt{Operator} at the given parameters, and (2) take a kernel and an \texttt{Operator}, but also the name of a concrete \texttt{ObjectiveFunction} subclass for custom pre- and post-processing around quantum circuit execution. Additionally, each of the public creation functions for \texttt{ObjectiveFunctions} requires the number of variational parameters in the quantum kernel. Optionally, programmers can provide a heterogeneous map of options that may affect the construction and use of the \texttt{ObjectiveFunction}.

Finally, the \texttt{Optimizer} concept represents a typical classical multi-variate function optimization strategy (COBYLA, L-BFGS, Adam, etc.). \texttt{Optimizers} expose an \texttt{optimize()} method that takes as input an \texttt{ObjectiveFunction}, which, as demonstrated above, is essentially a functor or lambda with the signature \texttt{double(const std::vector<double>, std::vector<double>\&)}. Here the first argument is the parameters to evaluate the \texttt{ObjectiveFunction} at, while the second argument represents the gradient vector as a reference that can be set. Using this functor signature, most classical derivative-free or gradient-based optimization routines are able to be implemented. As of this writing, \texttt{qcor} provides implementations of this interface that delegate to the popular NLOpt and MLPack libraries. 

\subsubsection{Task-based Asynchronous Execution}
The QCOR specification requires that implementations provide an optional asynchronous execution model for executing quantum-classical tasks (optional in the sense that one could still leverage synchronous execution if desired). Specifically, it defines a public API call, \texttt{taskInitiate()} which programmers invoke to launch a quantum-classical optimization task on a separate execution thread. Moreover, it defines a \texttt{Handle} type that is returned by \texttt{taskInitiate()} and is used by programmers to synchronize the host and execution thread (via a defined \texttt{sync(Handle\fix{\&})} call). The synchronization should cause the host thread to wait if the execution thread is not complete, and return a \texttt{ResultsBuffer} upon completion. The \texttt{ResultsBuffer} type is a simple data structure that provides the programmer with access to the optimal value and parameters. 

We implement this functionality in the QCOR runtime library implementation via the \texttt{std::future<T>} type provided by newer C\texttt{++} standards. Our implementation of \texttt{taskInitiate()} takes as input an \texttt{ObjectiveFunction} and an \texttt{Optimizer}, and returns a \texttt{Handle}, which is a typedef on \texttt{std::future<ResultsBuffer>}. The execution thread runs the \texttt{Optimizer} to compute the optimal parameters and value for the provided \texttt{ObjectiveFunction}. Programmers are free to do other work during execution of this asynchronous thread, and request the host and execution thread synchronize through the \texttt{sync(Handle\&)} call, returning a valid \texttt{ResultsBuffer} upon execution completion. The code snippet in Figure \ref{fig:task_code} demonstrates this workflow. 
\begin{figure}[h!] 
  \lstset {language=C++}
  \begin{lstlisting}
// Create the ObjectiveFunction
auto objective = createObjectiveFunction(
      ansatz, H, n_variational_params);

// Create the Optimizer.
auto optimizer = createOptimizer("nlopt");

// Launch the Optimization Task with taskInitiate
auto handle = taskInitiate(objective, optimizer);

// Go do other work...

// Query results when ready.
auto results = sync(handle);
printf("vqe-energy from taskInitiate = %f\n", 
                results.opt_val);
\end{lstlisting}
\caption{Demonstration of leveraging the \texttt{taskInitiate()} call and the \texttt{qcor} asynchronous execution model.}
\label{fig:task_code}
\end{figure}

\subsection{Compiler}
The \texttt{qcor} compiler implementation handles the complexity behind enabling this novel quantum-C++ language extension through simple extensions to Clang and integration with the QCOR runtime library. 
Here we go into detail behind the compiler implementation. We specifically highlight our novel implementation of the new Clang \texttt{SyntaxHandler} plugin, the overall compiler workflow, and the implementation of a compiler pass manager enabling general transformations on the compiled quantum kernel representation (for both optimization and placement). Ultimately, we put forward a \texttt{qcor} compiler executable that provides the same compiler flags programmers are used to, in addition to quantum-specific command line arguments. 

\subsubsection{Syntax Handler}
\label{sec:qcor_syntax}
The Clang compiler front-end exposes a modular and extensible set of libraries for common tasks found in the mapping of C, C\texttt{++}, and Objective-C source files to LLVM IR. It has a number of plugin interfaces, or extension points, that enable analysis of the abstract syntax tree (AST) representation of a C\texttt{++} source file. This extensibility enables a single Clang binary install to take on new functionality depending on what plugins are loaded at compile time via standard command line arguments. This approach is optimal for us and the \texttt{qcor} compiler implementation. We seek to enable quantum-classical programming in C\texttt{++} without having to modify core Clang/LLVM source bases, forcing a fork of these efforts and increasing the cost of maintainability for \texttt{qcor}. 

Our approach leverages a recent plugin interface contribution to Clang - the \texttt{SyntaxHandler} - which provides a hook for plugin developers to analyze functions written in any domain specific language (DSL) and provide a rewritten token stream to Clang that is composed of valid C\texttt{++} API calls (see Figure \ref{fig:clang_sh}). This replacement occurs after lexing and preprocessing, but before the AST is generated. This plugin interface exposes a \texttt{GetReplacement()} method that provides the function body tokens for implementation-specific analysis, and an output stream that the implementation uses to provide valid C\texttt{++} replacement code. The \texttt{SyntaxHandler} infrastructure will then replace the invalid DSL code with the provided output stream code and restart tokenization at the beginning of the function. Developers are free to update the function body, but can also write new code after it. Additionally, the \texttt{SyntaxHandler} exposes an \texttt{AddToPredefines()} method that can be used by implementations to add to the current source file's header file include statements. 

Our goal is to provide a \texttt{SyntaxHandler} implementation that enables the \texttt{qcor} C\texttt{++} language extension. Specifically, we want our users to be able to express quantum kernels in a quantum language agnostic manner, while retaining standard C\texttt{++} control flow statements and variable declaration and utility. To do so, we implement the \texttt{QCORSyntaxHandler} (see Figure \ref{fig:qcor_sh_uml}), with name \texttt{qcor}, which analyzes the incoming Clang \texttt{CachedTokens} reference and attempts to perform two tasks: (1) translate the quantum code itself into appropriate \texttt{QuantumRuntime} API calls, and (2) define a \texttt{QuantumKernel<Derived, Args...>} sub-type and associated function calls. 

\begin{figure}[b!] 
  \lstset {language=C++}
  \begin{lstlisting}
__qpu__ void bell(qreg q) {
  H(q[0]);
  using qcor::openqasm;
  cx q[0], q[1];
  using qcor::xasm;
  for (int i = 0; i < q.size(); i++) {
    Measure(q[i]);
  }
}
----- After Token Collection ----------
quantum::h(q[0]);
quantum::cx(q[0], q[1]);
for (int i = 0; i < q.size(); i++) {
  quantum::mz(q[i]);
}
\end{lstlisting}
\caption{Demonstration of mixing quantum languages within a quantum kernel, enabled via the \texttt{TokenCollector} infrastructure.}
\label{fig:sh_mix_lang}
\end{figure}
The first task relies on a further extension point called the \texttt{TokenCollector}, which we implement for the various quantum languages that we support. \texttt{qcor} currently has support (\texttt{TokenCollector} implementations) for XASM, OpenQasm, Quil, and a special circuit synthesis language that lets programmers describe their quantum code at the unitary matrix level. The \texttt{TokenCollector} exposes a single \texttt{collect()} method that allows implementations to map incoming clang \texttt{Tokens} to functional \texttt{QuantumRuntime} API calls dependent on the language corresponding to the implementation. Those \texttt{QuantumRuntime} calls are written to a provided \texttt{std::stringstream} that is passed down from the \texttt{QCORSyntaxHandler}. A unique feature of this architectural decomposition is that one can switch \texttt{TokenCollectors} while analyzing a given sequence of \texttt{CachedTokens}. This means that, dependent on some language extension syntax, one can define quantum kernels using multiple quantum languages within the same quantum kernel. In \texttt{qcor}, the default quantum kernel language is XASM, but we permit switching to other languages via a \texttt{using qcor::LANG;} statement. So to switch from the default XASM to OpenQasm for instance, and trigger internally a switch to the OpenQasm \texttt{TokenCollector}, one would simply write \texttt{using qcor::openqasm;} (see Figure \ref{fig:sh_mix_lang}). This is a useful feature since some languages do provide more efficient expressability for various quantum programming tasks. 

After the token collection phase of the \texttt{QCORSyntaxHandler} workflow, the provided \texttt{std::stringstream} contains the re-written \texttt{QuantumRuntime} API code for creating and executing the described quantum kernel. The details of how each \texttt{TokenCollector} implementation works is of critical importance. The most well-supported \texttt{TokenCollector} in \texttt{qcor} is the \texttt{XASMTokenCollector}. This implementation works by leveraging the XACC XASM \texttt{Compiler} implementation on a statement-by-statement basis. Specifically, it will attempt to compile each statement with this \texttt{Compiler} in order to map the statement to an XACC \texttt{Instruction} instance. If that mapping succeeds, the \texttt{Instruction} is mapped to a \texttt{QuantumRuntime} API call via an appropriate XACC \texttt{InstructionVisitor} (e.g. the \texttt{H(q[0])} call mapped to a \texttt{quantum::h(q[0])} call. If that mapping fails, the statement string itself is retained, and is assumed to be some classical code that must be part of the \texttt{QuantumRuntime} re-written source string (e.g. the \texttt{for} statement in Figure \ref{fig:sh_mix_lang}). The \texttt{OpenQasmTokenCollector} collects the incoming Clang \texttt{Tokens} and leverages the XACC Staq \texttt{Compiler} implementation to map each OpenQasm statement to an XACC \texttt{Instruction} instance. 

\begin{figure}[b!] 
  \lstset {language=C++}
  \begin{lstlisting}
__qpu__ void unitary(qreg q) {
  decompose {
    // Create the unitary matrix
    UnitaryMatrix ccnot_mat = 
            UnitaryMatrix::Identity(8, 8);
    ccnot_mat(6, 6) = 0.0;
    ccnot_mat(7, 7) = 0.0;
    ccnot_mat(6, 7) = 1.0;
    ccnot_mat(7, 6) = 1.0;
  }
  (q);
}

\end{lstlisting}
\caption{Demonstration of programming at the unitary matrix level using the \texttt{UnitaryMatrixTokenCollector}.}
\label{fig:decompose_unitary}
\end{figure}

We have also developed a means for programming at the unitary matrix level through an appropriate implementation of the \texttt{TokenCollector}. First, we define the \texttt{qcor::UnitaryMatrix} data structure, which is simply a \texttt{typedef} for a complex matrix provided by the \texttt{Eigen} matrix library \cite{eigen}. Next, we enable a \texttt{decompose} keyword as part of our quantum kernel language extension, which programmers declare, open a new scope, and define their unitary matrix using the \texttt{qcor::UnitaryMatrix} API. Programmers close that new scope and provide further arguments indicating the \texttt{qreg} to operate on, and information about the specific circuit synthesis algorithm to employ in decomposing the unitary matrix to gate-level quantum instructions. Figure \ref{fig:decompose_unitary} demonstrates how this circuit synthesis mechanism can be leveraged. Effectively, the \texttt{UnitaryMatrixTokenCollector} will be invoked when the \texttt{decompose} syntax is observed during token analysis, and will rewrite the kernel to delegate the decomposition of the unitary matrix to appropriate XACC circuit synthesis routines. 
\begin{figure}[t!] 
  \lstset {language=C++}
  \begin{lstlisting}
void bell(qreg q) {
  void __internal_call_function_bell(qreg);
  __internal_call_function_bell(q);
}
class bell : 
    public qcor::QuantumKernel<class bell_multi, 
                                    qreg> {
  friend class 
            qcor::QuantumKernel<class bell, qreg>;

protected:
  void operator()(qreg q) {
    if (!parent_kernel) {
      parent_kernel = 
        qcor::__internal__::
                    create_composite(kernel_name);
    }
    quantum::set_current_program(parent_kernel);
    quantum::h(q[0]);
    quantum::cnot(q[0], q[1]);
    for (int i = 0; i < q.size(); i++) {
      quantum::mz(q[i]);
    }
  }

public:
  inline static const std::string 
                kernel_name = "bell";
  bell(qreg q) : QuantumKernel<bell, qreg>(q) {}
  bell(std::shared_ptr<qcor::CompositeInstruction> 
                _parent, qreg q)
      : QuantumKernel<bell, qreg>(_parent, q) {}
  virtual ~bell() {
    auto [q] = args_tuple;
    operator()(q);
    if (is_callable) {
      quantum::submit(q.results());
    }
  }
};
void bell(
    std::shared_ptr<qcor::CompositeInstruction> 
                    parent, qreg q) {
  class bell_multi k(parent, q);
}
void __internal_call_function_bell(qreg q) {
  class bell_multi k(q);
}
\end{lstlisting}
\caption{The \texttt{QCORSyntaxHandler} translates quantum kernels (like the kernel in Figure \ref{fig:qcor_simple}) into new function calls and a \texttt{QuantumKernel<Derived,Args...>} subclass definition. }
\label{fig:clang_sh_code}
\end{figure}

The second task for the \texttt{QCORSyntaxHandler} is to rewrite the quantum kernel function and define a new \texttt{QuantumKernel<Derived, Args...>} sub-type, incorporating the results of the first task - the re-written \texttt{QuantumRuntime} code. Rewriting the function call as a \texttt{QuantumKernel} sub-type gives us auto-generated \texttt{adjoint / ctrl} methods, and provides an avenue for future kernel extensions enabling novel functionality. Our rewrite strategy is as follows: (1) rewrite the original function to forward declare a \texttt{\_\_internal\_call\_function\_KERNELNAME} function and immediately call that function (its implementation will follow the \texttt{QuantumKernel} sub-type declaration), (2) define the \texttt{QuantumKernel} subtype, and implement its \texttt{operator()(Args...)} method with the \texttt{QuantumRuntime} code generated from the token handling phase, (3) define the internal function call we forward declared in the original function, with an implementation that simply instantiates a temporary instance of the new \texttt{QuantumKernel} sub-type (immediately calling the destructor which affects quantum backend execution of the quantum code). An example of this re-write is given in Figure \ref{fig:clang_sh_code}. We also add a function after the sub-type definition that takes a \texttt{CompositeInstruction} as its first argument, which is used internally to enable kernel composition. 

% decompose unitary

Programmers see quantum kernel functions, but at compile time, these function are expanded into a new subclass definition of the \texttt{QuantumKernel}. The first subclass constructor takes as input the original function arguments, and calls the corresponding constructor on the superclass. This configures the kernel to be callable (\texttt{is\_callable = true;}). In the case of a \texttt{NISQ} \texttt{QuantumRuntime}, instantiation and destruction of a kernel constructed this way will build up the internal \texttt{CompositeInstruction} via the \texttt{QuantumRuntime} API calls, and invoke \texttt{submit()} to execute on the backend \texttt{Accelerator}. For the \texttt{FTQC} \texttt{QuantumRuntime}, instantiate and destruction invokes the \texttt{QuantumRuntime} calls which immediately affect execution of the single instruction on the backend \texttt{Accelerator}. Note that if the kernel has not been called, then the \texttt{\_parent\_kernel} is null, so the first task of \texttt{operator()(Args...)} is to create it. It is then given to the \texttt{QuantumRuntime} API and used for construction, or immediate execution, of the circuit. The second constructor takes as its first argument an already constructed \texttt{\_parent\_kernel}, which is set on the new instance's \texttt{\_parent\_kernel} attribute. Now when \texttt{operator()(Args...)} is called, a new \texttt{\_parent\_kernel} is not created, and the incoming one from instantiation is used. This directly enables kernel composition - the second constructor is always used for quantum kernels called from other quantum kernels. If this second constructor is used, then \texttt{is\_callable = false}, and \texttt{submit()} is never called on the kernel. For remote execution, submission to the backend is only ever invoked for entry-level quantum kernels. 

\subsubsection{Pass Manager}
\label{sec:Pass Manager}
As mentioned above, the \texttt{QuantumRuntime} API exposes a \texttt{submit()} call that affects execution of the constructed \texttt{CompositeInstruction} on the desired backend \texttt{Accelerator}. Upon invocation of this call, the runtime-resolved quantum IR tree is completely flattened and only contains simple quantum assembly instructions to be submitted to the specified QPU. Therefore, this submission API is ammenable for the implementation of a just-in-time (JIT) quantum circuit optimization and transformation sub-system which utilizes best-known techniques in the field of circuit optimization to further simplify the circuit before sending it to the target QPU. Since \texttt{qcor} is built upon the XACC framework, it is well-positioned to serve as an integration framework for state-of-the-art quantum compilation strategies coming from experts in the field. We specifically design our JIT quantum compilation system to build upon XACC's plugin extensibility in order to enable a diverse set of quantum compilation strategies.

Adopting the ubiquitous LLVM optimization framework pattern for user-contributed IR transformation strategies, we structure runtime circuit optimization tasks into \emph{passes} that simplify the input circuit in terms of gate count and depth. The application of runtime optimization passes is handled by a class called \texttt{PassManager}, and passes are implemented as subtypes of the XACC \texttt{IRTransformation}, and are invoked by the \texttt{PassManager}. This approach enables the \texttt{qcor} \texttt{PassManager} to inherit a well-established set of circuit optimizers from XACC, such as the implementations of the rotation folding and the phase polynomial optimization algorithms. Table \ref{tab:pass-desc} provides the default circuit optimizer passes (\texttt{xacc:IRTransformations}) that \texttt{qcor} leverages.

\begin{table}
\caption{Descriptions of circuit optimization passes that are implemented for \texttt{qcor}.}
\label{tab:pass-desc}
 \begin{tabular}{| p{0.15\textwidth} | p{0.33\textwidth} |} 
 \hline
 Pass Name & Description  \\ 
 \hline
 circuit-optimizer & A collection of simple pattern-matching-based circuit optimization routines.   \\ 
 \hline
 single-qubit-gate-merging & Combines adjacent single-qubit gates and finds a shorter equivalent sequence if possible.\\ 
 \hline
 two-qubit-block-merging & Combines a sequence of adjacent one and two-qubit gates operating on a pair of qubits and tries to find a more optimal gate sequence via Cartan decomposition if possible.\\ 
 \hline
 rotation-folding & A wrapper of the Staq's RotationOptimizer \cite{staq} which implemented the rotation gate merging algorithm. \\ 
 \hline
 voqc & A wrapper of the VOQC (Verified Optimizer for Quantum Circuits) OCaml library \cite{voqc}, which implements generic gate propagation and cancellation optimization strategy.   \\ 
 \hline
\end{tabular}
\end{table}

Based on internal profiling, we further define optimization \emph{levels} which dictate the set of passes and their execution order. The goal here is to strike a balance between the potential gate count reduction and the optimization time. For example, invoking the \texttt{qcor} compiler with ``\texttt{-opt 1}'' command-line option will activate optimization level 1. It is worth noting that since this option controls the final JIT optimization of the quantum kernel before remote execution, it will not impact the compile time of top-level classical-quantum code. The produced executable will contain the selected optimization level to pass over to the \texttt{PassManager} which then selects and loads appropriate \texttt{IRTransformation} modules to optimize the quantum IR tree. Once all passes have completed, the simplified circuit will be sent to the QPU for execution. 

More advanced users can also specify an ordered list of passes to be executed by using the \texttt{qcor}'s ``\texttt{-opt-pass}'' option. External developers can thus develop in-house passes adhering to the \texttt{IRTransformation} API and integrate them into the \texttt{qcor} compilation and execution workflow using this compile option. For example, we have made available two \texttt{IRTransformation} plugins which wrap the C\texttt{++} Staq rotation folding \cite{staq} and the OCaml-based Verified Optimizer for Quantum Circuits (VOQC) \cite{voqc} optimizers, thereby demonstrating the cross-language extensibility of the \texttt{qcor} circuit optimization sub-system.

For diagnostic purposes, the \texttt{PassManager} analyzes detailed statistics about each pass, such as the execution time, the gate count distribution before and after the pass, which could be retrieved for analysis. In Section~\ref{sec:Optimizing Compiler Performance}, we will show some statistics of the passes that are currently available in the \texttt{qcor}-XACC ecosystem.  

\subsubsection{Placement}
\label{sec:Placement}
\begin{figure}[b!] 
  \lstset {language=C++}
  \begin{lstlisting}
// Create a multi-qubit entangled state 
__qpu__ void entangleQubits(qreg q) {
  H(q[0]);
  for (int i = 1; i < q.size(); i++) {
    CX(q[0],q[i]);
  }
  for (int i = 0; i < q.size(); i++) {
    Measure(q[i]);
  }
}

int main() {
  // Create a 4-qubit register
  auto q = qalloc(4);
  // Execute the kernel
  entangleQubits(q);
  // Expect: ~50-50 for "0000" and "1111"
  q.print();
}
  \end{lstlisting}
  \begin{lstlisting}
// Target ibmq_ourense backend:
// qcor -qpu aer:ibmq_ourense 
H q0             ------------------------
CNOT q1,q0       | Ourense Connectivity |
CNOT q0,q1       |  (0) -- (1) -- (2)   |
CNOT q1,q2       |          |           |
CNOT q1,q3       |         (3)          |
Measure q1       |          |           |
Measure q0       |         (4)          |
Measure q2       ------------------------
Measure q3
  \end{lstlisting}
\begin{lstlisting}  
// Target ibmqx2 (ibmq_5_yorktown) backend
// qcor -qpu aer:ibmqx2
H q0             -----------------------
CNOT q0,q1       | ibmqx2 Connectivity |       
CNOT q2,q0       |        (1)          |
CNOT q0,q2       |      /  |           |
CNOT q2,q3       |    (0)-(2)-(3)      |
Measure q2       |         | /         |
Measure q1       |        (4)          |
Measure q0       -----------------------
Measure q3
\end{lstlisting}
\caption{Code snippet demonstrating \texttt{qcor} placement. (Top) \texttt{qcor} source code and final circuits after placement for the IBMQ's (Middle) Ourense and (Bottom) Yorktown backends.}
\label{fig:qcor_placement_example}
\end{figure}
When \texttt{qcor} compiles the executable for a target accelerator backend, it also takes into account the qubit connectivity as well as any user-defined mappings to project the logical qubit indices as defined in the quantum kernel onto the actual physical qubit indices on hardware. This hardware placement functionality often involves (1) permutations of gates and qubits, e.g., by inserting SWAP gates, so that the resulting circuit satisfies the device topology constraints and (2) direct logical-physical qubit mapping to take advantage of best-performing qubits.

\begin{table}[!b]
\caption{Descriptions of hardware placement strategies that are implemented for QCOR.}
\label{tab:placement-desc}
 \begin{tabular}{| p{0.15\textwidth} | p{0.33\textwidth} |} 
 \hline
 Name & Description  \\ 
 \hline
 Swap shortest path & Implement permutation-based mapping for uncoupled qubits~\cite{staq}.\\ 
 \hline
 Noise Adaptive & Optimize a noise-adaptive layout~\cite{murali2019noise} based on backend calibration data (gate errors.)\\ 
 \hline
 Sabre & Implement SWAP-based BidiREctional heuristic search algorithm (SABRE)~\cite{li2019tackling}.\\ 
 \hline
 QX Mapping & Implement the IBM-QX contest-winning technique~\cite{zulehner2018efficient}. \\ 
 \hline
\end{tabular}
\end{table}

To address the first task, \texttt{qcor} defaults to an \texttt{xacc::IRTransformation} implementation delegating to the Staq \cite{staq} library providing a generic shortest path permutation algorithm (\texttt{swap-shortest-path}) whereby two-qubit gates between uncoupled qubits are swapped to satisfy the coupling graph. Figure~\ref{fig:qcor_placement_example} demonstrates such mapping when we compile the same kernel source for two different IBMQ device targets, namely the Ourense and Yorktown 5-qubit backends. 
Since their connectivity graphs are different, the resulting circuits after placement are also different. Specifically, the sequence of CNOT gates was permutated to match the backend topology and the measure gates are also swapped accordingly. It is worth noting that this propagating permutation approach is more efficient than a SWAP gate-based solution since we do not need to swap the qubits back and forth.
Besides \texttt{swap-shortest-path}, Table~\ref{tab:placement-desc} provides the details of hardware placement strategies that are available in \texttt{qcor}.

Manual qubit-to-qubit mapping functionality is also available in \texttt{qcor}. In particular, by supplying a `\texttt{-qubit-map}' option along with a sequence of qubit indices to \texttt{qcor}, the runtime placement service will map logical qubits to the physical ones according to this map. For example, depending on the readout and gate error information of the backend, we may want to use qubit 5 and 6 for a two-qubit quantum kernel which was written in terms of \texttt{q[0]} and \texttt{q[1]} by simply compiling with `\texttt{-qubit-map 5,6}`.

\subsubsection{Automated Error Mitigation}
\begin{figure}[b!] 
  \lstset {language=C++}
  \begin{lstlisting}
__qpu__ void noisy_zero(qreg q) {
    for (int i = 0; i < 100; i++) {
        X(q[0]);
    }
    Measure(q[0]);
}

int main() {
    qreg q = qalloc(1);
    noisy_zero(q);
    std::cout << "Expectation: " 
                << q.exp_val_z() << "\n";
}
-------------------------------------------------
$ qcor -qpu aer[noise-model:noise_model.json] \
    -shots 4096 -o noisy.x zne_test.cpp 
$ ./noisy.x
Expectation: 0.895996
$ qcor -qpu aer[noise-model:noise_model.json] \ 
    -shots 4096 -em mitiq -o mitiq_noise.x \ 
    zne_test.cpp 
$ ./mitiq_noise.x
Expectation: 1.02295
\end{lstlisting}
\caption{Code snippet demonstrating QCOR automated error mitigation leveraging the Mitiq library, specifically zero-noise extrapolation.}
\label{fig:noisy_em}
\end{figure}
XACC enables automated error mitigation via decoration of the \texttt{Accelerator} backend \cite{mccaskey2020xacc}. The \texttt{AcceleratorDecorator} service interface inherits from \texttt{Accelerator} but also contains an \texttt{Accelerator} member reference, enabling an \texttt{Accelerator::execute()} override that provides an opportunity for pre- and post-processing around execution of the decorated \texttt{Accelerator}. For error mitigation, this is used to analyze or update the incoming compiled circuit, execute it, and analyze and mitigate the results based on the sub-type's implemented strategy. Since \texttt{qcor} builds upon XACC and ultimately targets backend \texttt{Accelerators}, this mechanism should also be readily available to users of the \texttt{qcor} language extension and compiler. 

We have added this capability to \texttt{qcor} via an \texttt{-em} command line option. This option flag lets users specify the name of a decorator to use to automatically apply error mitigation to kernel invocations. The code snippet in Figure \ref{fig:noisy_em} demonstrates this, whereby we have a quantum kernel that applies a large, even number of \texttt{X} gates on a single qubit, theoretically resulting in the $\ket{0}$ state. Due to the presence of noise, this will not be the case, and we should observe an expectation value with respect to \texttt{Z} measurements that drifts from the true value of 1.0. The bottom half of this snippet shows how one would use this error mitigation flag. Here we compile to the IBM noise-aware \texttt{Aer} simulation backend, providing a custom noise model file as an option. Execution of this compiled executable results in a noisy expectation value, as expected. We next compile with the same noise model but additionally indicate we'd like to apply error mitigation from the Mitiq library \cite{mitiq, mitit-paper}, which provides routines for zero-noise extrapolation \cite{mitiq_zne}. Executing the compiled executable this time we see that the result has been shifted closer to the true value of 1.0. \texttt{qcor} enables one to stack these decorators by passing more than one \texttt{-em} flag, and the order with which they are seen on the command line will represent the order they will be executed. 

\subsubsection{Compiler Workflow}
The architecture described above ultimately puts forward C\texttt{++} libraries that provide pertinent \texttt{qcor} runtime and compile-time capabilities. In order for programmers to interface with this novel infrastructure, we provide a \texttt{qcor} compiler command-line executable. This executable is meant to directly mimic existing compilers like \texttt{clang++} and \texttt{g++}, but with the addition of quantum-pertinent command line options. We provide this compiler as an executable Python script, which delegates to a \texttt{clang++} sub-process call configured with all necessary include paths, library link paths, libraries, and compiler flags required for executing the \texttt{qcor} compilation workflow. Of critical importance is the loading of the \texttt{QCORSyntaxHandler} plugin library, which enables the underlying \texttt{clang++} call to operate on defined quantum kernels and transform them to valid C\texttt{++} code. Additionally, the \texttt{qcor} compiler exposes a \texttt{-qpu} compiler flag that lets users dictate what quantum backend this source file should be compiled to. The quantum backend name provided follows the XACC syntax for specifying \texttt{Accelerators} (e.g. \emph{accelerator\_name:backend\_name}). As seen in Section \ref{sec:Pass Manager}, the compiler also exposes \texttt{-opt LEVEL} and \texttt{-opt-pass PASSNAME} arguments to turn on quantum circuit optimization. Just like existing classical compilers, \texttt{qcor} can be used in compile-only mode (\texttt{-c SOURCEFILE.cpp}) as well as in link-mode. 

The overall compiler workflow is fairly simple, and can be described as follows: (1) invocation of \texttt{qcor} on a quantum-classical C\texttt{++} source file, indicating the backend QPU to target, (2) \texttt{clang++} is invoked and loads the \texttt{QCORSyntaxHandler} plugin library, (3) usual Clang preprocessing and lexing occurs, (4) the \texttt{QCORSyntaxHandler} is invoked on all \texttt{\_\_qpu\_\_} annotated functions, translating them to a set of new functions and a \texttt{QuantumKernel} definition, as in Figure \ref{fig:clang_sh_code}, and (5) finally, classical compilation proceeds with this rewrite (AST generated, LLVM IR CodeGen executed). The user is left with a classical binary executable or object file (depending on whether \texttt{-c} was used). Invocation of the executable proceeds as it would normally (\texttt{./a.out}, or whatever the executable was named). 

\begin{figure}[b!] 
  \lstset {language=C++}
  \begin{lstlisting}
#include "qcor_jit.hpp"
int main() {

  // QJIT is the entry point to QCOR quantum kernel 
  // just in time compilation
  QJIT qjit;

  // Define a quantum kernel string dynamically
  const auto kernel_src = R"#(
    __qpu__ void bell(qreg q) {
        using qcor::openqasm;
        h q[0];
        cx q[0], q[1];
        creg c[2];
        measure q -> c;
    })#";

  // Use qjit to compile this at runtime
  qjit.jit_compile(kernel_src);

  // Now, one can get the compiled kernel as a 
  // functor to execute, must provide the kernel 
  // argument types as template parameters
  auto bell = qjit.get_kernel<qreg>("bell");

  // Allocate a qreg and run the kernel functor
  auto q = qalloc(2);
  bell(q);
  q.print();

  // Or, one can call the QJIT invoke method 
  // with the name of the kernel function and 
  // the necessary function arguments.
  auto r = qalloc(2);
  qjit.invoke("bell", r);
  r.print();
}
\end{lstlisting}
\caption{Code snippet demonstrating QCOR quantum kernel just in time compilation.}
\label{fig:qjit}
\end{figure}
\subsubsection{Just-in-Time Quantum Kernel Compilation}
Another architectural point of note for the compiler is the addition of data structures and utilities to perform just-in-time compilation of quantum kernels. We foresee use cases whereby developers may wish to build up quantum circuits at runtime based on pertinent runtime information. This is difficult with quantum kernel function declarations, as these are defined at compile time. We have therefore introduced a new data type, \texttt{QJIT}, which provides quantum kernel just-in-time execution (JIT). The code snippet in Figure \ref{fig:qjit} demonstrates how one might use this utility. \texttt{QJIT} exposes a \texttt{jit\_compile()} method that takes as input the quantum kernel as a source string. This method will then programmatically run the \texttt{QCORSyntaxHandler} on that source string to produce the source string containing the \texttt{QuantumKernel} sub-type definition plus additional utility functions (as in Figure \ref{fig:clang_sh_code}). This new source string is then compiled to an LLVM IR \texttt{Module} instance using the Clang \texttt{CodeGenAction} programmatically. The resultant \texttt{Module} is then passed to the LLVM JIT utility data structures (\texttt{ExecutionSession}, \texttt{IRCompileLayer}) for just-in-time compilation. Finally, a pointer to the representative function for the quantum kernel is stored and returned via the \texttt{QJIT::get\_kernel<Args...>()} call, or leveraged in the \texttt{QJIT::invoke()} call. In this way, programmers can compile source string dynamically at runtime, and get a function pointer reference to the JIT compiled function for future execution. This workflow also incorporates \texttt{Module} caching so that the same quantum kernel source code is not re-compiled every time it is encountered (or the executable running this workflow is run).

\section{Demonstration}
Now we turn to some illustrative examples of using the \texttt{qcor} compiler infrastructure. Specifically, we detail code snippets demonstrating the level of quantum-classical programmability that \texttt{qcor} provides, as well as novel library data structures and API calls for affecting execution of useful quantum algorithms (VQE \cite{VQE} QAOA \cite{adapt_qaoa}, QPE \cite{qpe}, etc). 
\subsection{Quantum Phase Estimation}
\begin{figure}[t!] 
  \lstset {language=C++}
  \begin{lstlisting}
// QCOR standard libraries
#include "qft.hpp"

// The Oracle: a T gate
__qpu__ void compositeOracle(qreg q) {
  // T gate on the last qubit
  int last_qbit = q.size() - 1;
  T(q[last_qbit]);
}

// Main algorithm
__qpu__ void QuantumPhaseEstimation(qreg q) {
  const auto nQubits = q.size();
  // Prepare eigenstate (|1>)
  X(q[nQubits - 1]);

  // Apply Hadamard gates to the counting qubits:
  for (int qIdx = 0; qIdx < nQubits - 1; ++qIdx) {
    H(q[qIdx]);
  }

  // Apply Controlled-Oracle
  const auto bitPrecision = nQubits - 1;
  for (int32_t i = 0; i < bitPrecision; ++i) {
    const int nbCalls = 1 << i;
    for (int j = 0; j < nbCalls; ++j) {
      int ctlBit = i;
      // Controlled-Oracle:
      // in this example, Oracle is T gate;
      // i.e. Ctrl(T) = CPhase(pi/4)
      compositeOracle::ctrl(ctlBit, q);
    }
  }

  // Inverse QFT on the counting qubits:
  int startIdx = 0;
  int shouldSwap = 1;
  iqft(q, startIdx, bitPrecision, shouldSwap);

  // Measure counting qubits
  for (int qIdx = 0; qIdx < bitPrecision; ++qIdx) {
    Measure(q[qIdx]);
  }
}

// Executable entry point:
int main(int argc, char **argv) {
  // Allocate 4 qubits, i.e. 3-bit precision
  auto q = qalloc(4);
  QuantumPhaseEstimation(q);
  // dump the results
  // EXPECTED: only "100" bitstring
  q.print();
}
\end{lstlisting}
\caption{Code snippet demonstrating the Quantum Phase Estimation algorithm.}%. Compile and run on a simulator with \texttt{qcor -qpu qpp -shots 1024 qpe.cpp ; ./a.out}}
\label{fig:qcor_qpe_example}
\end{figure}
The quantum phase estimation (QPE) algorithm is a seminal quantum subroutine that computes the eigenvalue of a unitary matrix for a given eigenvector. 
From a programming perspective, this algorithm demonstrates some intriguing aspects of the composability and synthesis of quantum programs. In particular, the input to the algorithm is a black box operation $U$ (\emph{oracle}) which we must be able to apply conditioned on a qubit. Hence, the compiler needs to figure out the decomposition in terms of basic gates to implement that arbitrary controlled-U operation. In \texttt{qcor}, each user-defined quantum kernel has intrinsic \texttt{adjoint()} and \texttt{ctrl()} extensions, which automatically generate the adjoint and controlled circuits. 
%This functionality is useful in constructing the QPE algorithm, as shown in Figure~\ref{fig:qcor_qpe_example}.

We demonstrate the programmability of the QPE algorithm in Figure \ref{fig:qcor_qpe_example}. The oracle is expressed as a \texttt{qcor} kernel (annotated with \texttt{\_\_qpu\_\_}) named \texttt{compositeOracle} which only contains a single T gate operating on the last qubit of the provided quantum register. It is worth noting that the oracle can be an arbitrarily complex circuit or even be specified as a unitary matrix using the \texttt{qcor} unitary decompose extension. 
Given this oracle kernel, the QPE algorithm requires the application of controlled-$U^k$ operations. Thanks to the ubiquitous \texttt{for} loop and the built-in \texttt{ctrl} kernel extension, the algorithm is expressed in a very succinct manner yet generic for arbitrary oracles.

There is another language feature that we also want to point out in this example. We take advantage of the Inverse Quantum Fourier Transform (\texttt{iqft}) kernel that is pre-defined in the \texttt{qcor} standard libraries by simply including the appropriate header file (\texttt{qft.hpp}). The algorithm is implemented for generic cases allowing us to specify a subset of the qubit register to act upon and to control whether or not we need to add SWAP gates at the beginning of the circuit.

\subsection{GHZ State on a Physical Backend}
\begin{figure}[b!] 
  \lstset {language=C++}
  \begin{lstlisting}
__qpu__ void ghz(qreg q) {
    H(q[0]);
    for (int i = 0; i < q.size()-1; i++) {
        CX(q[i], q[i+1]);
    }
    for (int i = 0; i < q.size(); i++) {
        Measure(q[i]);
    }
}
// helper to show histogram of counts
void plot_counts(auto&& counts) {...}
int main() {
    auto q = qalloc(5);
    ghz::print_kernel(std::cout, q);
    ghz(q);
    plot_counts(q.counts());
}
\end{lstlisting}
\caption{Code snippet demonstrating preparing a GHZ state on the 5 qubit \texttt{ibmq\_vigo} physical backend.}
\label{fig:ghz}
\end{figure}
\begin{figure}[h!] 

  \lstset {language=C++}
    % \begin{mdframed}[backgroundcolor=black!5,hidealllines=true]

  \begin{lstlisting}[escapeinside={(*@}{@*)}]
$ qcor -qpu ibm:ibmq_vigo ghz.cpp ; ./a.out
H q0            ---------------------
CNOT q0,q1      | Vigo Connectivity |
CNOT q1,q2      | (0) -- (1) -- (2) |
CNOT q2,q1      |         |         |
CNOT q1,q2      |        (3)        |
CNOT q2,q1      |         |         |
CNOT q1,q3      |        (4)        |
CNOT q3,q4      ---------------------
Measure q0
Measure q2
Measure q1
Measure q3
Measure q4
\end{lstlisting}
\caption{Standard out from code in Figure \ref{fig:ghz}. The default placement strategy has been applied to enable all CNOTs in the logical program.}
\label{fig:ghzr_out}
% \end{mdframed}
\end{figure}
\begin{figure}[t!] 
\includegraphics[width=.5\textwidth]{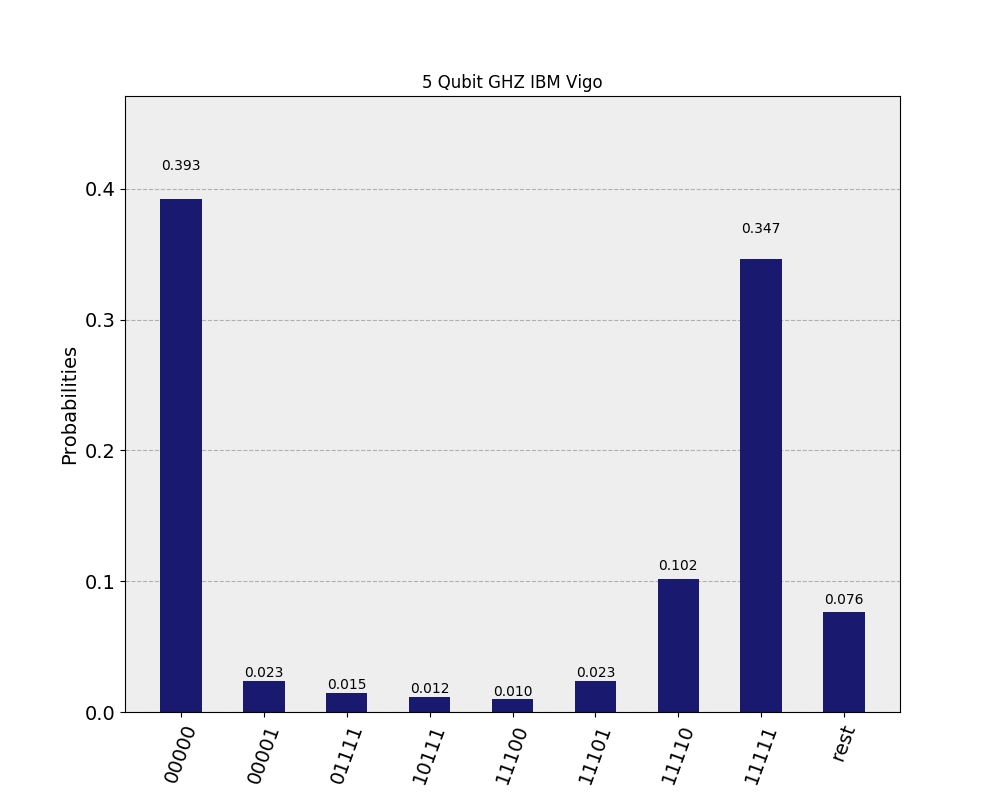}
\caption{Results of running the code in Figure \ref{fig:ghz} on the \texttt{ibm\_vigo} physical backend. Execution on Aug. 27, 2020. 11:03 AM EDT, IBM Job-Id: 5f47cb10654c28001b53b144}
\label{fig:ghzr}
\end{figure}

% \begin{figure}[h!]
% \centering  
% \includegraphics[width=.5\textwidth]{figures/ghz_vigo_results.png}
% \caption{Results of compiling and running the code in Figure \ref{fig:ghz} on the \texttt{ibm\_vigo} physical backend. Execution on Aug. 27, 2020. 11:03 AM EDT, IBM Job-Id: 5f47cb10654c28001b53b144}
% \label{fig:ghzr}
% \end{figure}
To demonstrate \texttt{qcor}'s ability to compile to physical backends, here we demonstrate a simple GHZ experiment on a 5-qubit physical backend from IBM. The logical connectivity of this problem will not directly map to the physical connectivity of the backend we target (\texttt{ibmq\_vigo}), but \texttt{qcor} handles this by applying an appropriate placement strategy, as described in Section \ref{sec:Pass Manager}. The code snippet in Figure \ref{fig:ghz} shows a simple kernel that runs the GHZ state on 5 qubits. In \texttt{main()}, we allocate the 5-qubit \texttt{qreg}, print the kernel in order to see the results of placement on the \texttt{ibm\_vigo} backend, run the kernel, and output the bit strings and corresponding counts observed. The results of compilation and execution of this code are shown in Figures \ref{fig:ghzr_out},\ref{fig:ghzr}, where one can clearly see the presence of the \texttt{SWAP} to enforce the logical connectivity of the program (introduced by the default Staq \texttt{swap-shortest-path} placement strategy). The results indicate the typical noise present in execution on NISQ hardware, but one can see the dominant observed configurations of \texttt{00000} and \texttt{11111}, as expected.

\subsection{Feed-Forward Error Correction}
\label{sec:Feed-Forward Error Correction}
\begin{figure}[t!] 
  \lstset {language=C++}
  \begin{lstlisting}
// Measure Z0Z1 and Z1Z2 syndromes 
// and recover from a bit-flip error.
__qpu__ void correctLogicalQubit(qreg q, 
                                int logicalIdx, 
                                int ancIdx) {
  int physicalIdx = logicalIdx * 3;
  // Step 1: Measure Z0Z1
  CX(q[physicalIdx], q[ancIdx]);
  CX(q[physicalIdx + 1], q[ancIdx]);
  // Measure the ancilla to determine the syndrome.
  const bool parity01 = Measure(q[ancIdx]);
  if (parity01) {
    // Reset ancilla qubit for reuse
    X(q[ancIdx]);
  }
  // Step 2: Measure Z1Z2
  CX(q[physicalIdx + 1], q[ancIdx]);
  CX(q[physicalIdx + 2], q[ancIdx]);
  // Measure the ancilla to determine the syndrome.
  const bool parity12 = Measure(q[ancIdx]);
  if (parity12) {
    // Reset ancilla qubit for reuse
    X(q[ancIdx]);
  }
  // Step 3: Correct bit-flip errors 
  // based on parity results:
  //     Error | Z0Z1 | Z1Z2
  //     ===================
  //       Id  |False |False
  //       X0  |True  |False
  //       X1  |True  |True
  //       X2  |False |True
  if (parity01 && !parity12) {X(q[physicalIdx]);}  
  if (parity01 && parity12) {X(q[physicalIdx+1]);}  
  if (!parity01 && parity12) {X(q[physicalIdx+2]);}
}

// Run a full QEC cycle on bit-flip code encoded 
// qubit register.
__qpu__ void runQecCycle(qreg q) {
  int nbLogicalQubits = q.size() / 3;
  int ancBitIdx = q.size() - 1;
  for (int i = 0; i < nbLogicalQubits; ++i) {
    correctLogicalQubit(q, i, ancBitIdx);
  }
}
\end{lstlisting}
\caption{Code snippet demonstrating bit-flip quantum error correction code. The \texttt{runQecCycle} kernel iterates over all \emph{logical} qubits (encoded as three consecutive physical qubits) and performs syndrome detection and correction (using \texttt{correctLogicalQubit} helper kernel). Compilation requires the \texttt{-qrt ftqc} flag.}
\label{fig:qkernel_ftqc}
\end{figure}

In this demonstration, we seek to illustrate the utility of the FTQC runtime to implement quantum error correction (QEC), which is a crucial aspect of fault-tolerant quantum computation. Specifically, we examine the implementation of the canonical QEC feedback (syndrome) and feed-forward (correction) loop of a toy three-qubit bit-flip encoding scheme, as shown in Figure~\ref{fig:qkernel_ftqc}.
The syndrome signatures (\texttt{parity01} and \texttt{parity12} boolean variables) detected by measurement operations are used to infer the most probable bit-flip location for correction.  Albeit its simplicity, this model of error correction immediately generalizes to other codes which could require much more complex decoding mechanisms such as the Blossom~\cite{fowler2015minimum} or maximum-likelihood ~\cite{bravyi2014efficient} algorithms for the surface code~\cite{fowler2012surface}.

\subsection{Multi-Language Kernel Development}
\begin{figure}[t!] 
  \lstset {language=C++}
  \begin{lstlisting}
__qpu__ void ccnot(qreg q, 
            std::vector<int> bit_config) {
  // Setup the initial bit configuration
  // This is using XASM language
  for (auto [i, bit] : enumerate(bit_config)) {
    if (bit) {
      X(q[i]);
    }
  }

  // Use the Unitary Matrix DSL for 
  // creating the Toffoli matrix to decompose
  decompose {
    UnitaryMatrix ccnot_mat = 
                UnitaryMatrix::Identity(8, 8);
    ccnot_mat(6, 6) = 0.0;
    ccnot_mat(7, 7) = 0.0;
    ccnot_mat(6, 7) = 1.0;
    ccnot_mat(7, 6) = 1.0;
  }(q);

  // Switch to OpenQasm and Measure all
  using qcor::openqasm;
  creg c[3];
  measure q -> c;
}

// Helper functions
std::vector<std::vector<int>> 
            generate(int size) {...}
void print_result(auto& bit_config, 
            auto counts) {...}

int main() {
  // Loop over all configs and print out 
  // the Toffoli truth table
  for (auto &bit_config : generate(3)) {
    auto q = qalloc(3);
    ccnot(q, bit_config);
    auto counts = q.counts();
    print_result(bit_config, counts);
  }
}
----------- compile and run with -------------
$ qcor -qpu qpp -shots 1024 ccnot.cpp && ./a.out
000 -> 000
001 -> 001
010 -> 010 
011 -> 011 
100 -> 100
101 -> 101
110 -> 111
111 -> 110
\end{lstlisting}
\caption{Code snippet demonstrating the mixing of available quantum languages via the \texttt{qcor} \texttt{TokenCollector} architecture.}
\label{fig:qcor_multi_lang_code}
\end{figure}
The \texttt{SyntaxHandler} and \texttt{TokenCollector} architecture gives us a unique opportunity for general embedded domain-specific language processing in C\texttt{++} for quantum programming. Moreover, as implemented, it gives us the ability to program kernels in multiple \texttt{qcor}-supported quantum languages. Here we demonstrate this capability using an example that leverages both gate-level and unitary matrix-level programming approaches side-by-side. 

Figure \ref{fig:qcor_multi_lang_code} demonstrates the generation of the truth table for the Toffoli gate using three distinct languages in a single quantum kernel definition. The example starts off by defining a controlled-CNOT quantum kernel (\texttt{ccnot}) that takes a \texttt{qreg} and a \texttt{vector<int>} describing the initial qubit state configuration (some combination of 0s and 1s). The kernel starts by using the XASM language to operate \texttt{X} gates on qubits with a corresponding bit configuration of 1 in the \texttt{bit\_config} vector. Next, the kernel leverages the unitary matrix decomposition DSL for describing the Toffoli interaction as a matrix. This tells \texttt{qcor} to decompose the corresponding unitary matrix with an internal circuit synthesis algorithm (QFAST \cite{qfast} by default). Finally, the kernel uses the OpenQasm language to apply measure gates to all qubits in the \texttt{qreg}. The \texttt{main()} implementation loops over all bit configurations, each time allocating a three-qubit \texttt{qreg}, executing the kernel, and printing the resultant truth table entry. 

\subsection{Incorporating Pre-Existing OpenQasm Codes}
\begin{figure}[t!] 
  \lstset {language=C++}
  \begin{lstlisting}
------------------- grover.qasm -------------------
OPENQASM 2.0;
include "qelib1.inc";
qreg qubits[9];
creg c[9];
x qubits[5];
h qubits[0];
h qubits[1];
ccx qubits[0],qubits[1],qubits[6];
ccx qubits[2],qubits[6],qubits[7];
ccx qubits[3],qubits[7],qubits[8];
... missing for brevity, file has 164 lines
h qubits[3];
h qubits[4];
------------------- grover.cpp --------------------
__qpu__ void grover(qreg q) {
   using qcor::openqasm;
   #include "grover.qasm"
   using qcor::xasm;
   for (int i = 0; i < q.size(); i++) {
     Measure(q[i]);
   }
}
int main() {
  auto q = qalloc(9);
  grover::print_kernel(std::cout, q);
  grover(q);
}
\end{lstlisting}
\caption{Code snippet demonstrating the inclusion of pre-existing OpenQasm files into quantum kernel expressions.}
\label{fig:grover_oq}
\end{figure}
A large number of benchmarks and application-level quantum codes are written as stand-alone OpenQasm files - standard text files containing OpenQasm quantum code. Integration of these pre-existing codes with the \texttt{qcor} quantum kernel expression mechanism is straightforward, and we demonstrate it here. The top part of the code snippet in Figure \ref{fig:grover_oq} shows the contents of an OpenQasm file called \texttt{grover.qasm}. The bottom part demonstrates a \texttt{qcor} C\texttt{++} file that incorporates this OpenQasm code into the usual quantum kernel function definition. Programmers simply note that the kernel language to be used is OpenQasm via the \texttt{using qcor::openqasm} statement, and then leverage the existing C\texttt{++} preprocessor to include the contents of the \texttt{grover.qasm} file within the function body. One can then add any other kernel code using any of the available kernel languages (e.g. adding measurements using XASM as seen in the code snippet). Programmers can then invoke the kernel on an appropriately sized \texttt{qreg} instance, or print the kernel qasm to see that the OpenQasm was appropriately incorporated. 

\subsection{Variational Algorithms with the QCOR API}
\begin{figure}[b!] 
  \lstset {language=C++}
  \begin{lstlisting}
__qpu__ void ansatz(qreg q, double theta) {
  X(q[0]);
  Ry(q[1], theta);
  CX(q[1], q[0]);
}

int main(int argc, char **argv) {
  // Create the Deuteron Hamiltonian
  auto H = 5.907 - 2.1433 * X(0) * X(1) 
           - 2.1433 * Y(0) * Y(1) + .21829 * Z(0)
           - 6.125 * Z(1);

  // Create the ObjectiveFunction
  auto objective = 
        createObjectiveFunction(ansatz, H, 1);

  // Create the Optimizer
  auto optimizer = createOptimizer("nlopt");

  // Call taskInitiate, kick off optimization 
  // of the give functor dependent on the 
  // ObjectiveFunction, async call
  auto handle = taskInitiate(objective, optimizer);

  // Go do other work...

  // Query results when ready.
  auto results = sync(handle);

  // Print the optimal value.
  printf("<H> = %f\n", results.opt_val);
}
-------------- compile/run with ---------------- 
$ qcor -qpu qpp qcor_api_example.cpp
$ ./a.out
\end{lstlisting}
\caption{Code snippet demonstrating the low-level \texttt{qcor} API for variational tasks.}
\label{fig:qcor_var_spec}
\end{figure}
Here we demonstrate the utility of the public API and data structures defined by the QCOR specification, and specifically its application to hybrid variational algorithms. The code snippet in Figure \ref{fig:qcor_var_spec} provides an example of computing the ground state energy of the two qubit deuteron Hamiltonian using the \texttt{qcor} \texttt{Operator}, \texttt{ObjectiveFunction}, \texttt{Optimizer}, and \texttt{taskInitiate()}. The example starts with a quantum kernel definition describing the variational quantum circuit, in this case a simple kernel leveraging the XASM language using a single \texttt{double} parameter. \texttt{main()} begins with the definition of the \texttt{Operator} describing the Hamiltonian for this system, which is extremely natural when leveraging the \texttt{qcor} \texttt{X}, \texttt{Y}, \texttt{Z} function calls. Next, the programmer creates an \texttt{ObjectiveFunction}, giving it the quantum kernel, \texttt{Operator}, and the number of variational parameters in the problem. Note that when one does not provide the name of the \texttt{ObjectiveFunction} sub-type, \texttt{vqe} is assumed. Next, the \texttt{Optimizer} is created, specifically an implementation backed by the NLOpt library, defaulting to the COBYLA derivative-free algorithm. The optimization task is launched on a separate execution thread via the \texttt{taskInitiate()} call, returning a \texttt{Handle} which is kept and used later to synchronize the host and execution threads. Finally, after synchronization, the optimal value can be retrieved from the \texttt{ResultsBuffer}.

\subsection{Overall Compiler Performance}
\label{sec:Optimizing Compiler Performance}
Here we demonstrate the overall effectiveness of \texttt{qcor} as an quantum compiler. To start, we demonstrate the performance of our JIT circuit optimization procedure (described in Section \ref{sec:Pass Manager}) by running the \texttt{qcor} compiler with flag \texttt{-opt 1} on a collection of common benchmark circuit files. We compare our optimization passes to existing approaches from the Staq compiler executable. Since we have also wrapped the Staq rotation-folding optimization as a pass that \texttt{qcor} can use (an \texttt{xacc::IRTransformation}), we are able to directly compare the performance between passes. More importantly, as mentioned in Section~\ref{sec:Pass Manager}, we have bundled those passes into a custom level, which instructs the \texttt{PassManager} to execute passes in series. In particular, the overall level-1 optimization performance in Table \ref{tab:opt-data} is the result of the \texttt{rotation-folding}, \texttt{single-qubit-gate-merging}, \texttt{circuit-optimizer}, and \texttt{voqc} (see Table~\ref{tab:pass-desc} for descriptions) sequence.
\begin{table}[!h]
\caption{Circuit optimization results for \cite{staq} benchmarks using (1) individual \texttt{qcor} passes and (2) \texttt{qcor}'s level-1 optimization sequence.}
\label{tab:opt-data}
 \begin{tabular}{ p{0.23\textwidth}  p{0.07\textwidth} p{0.07\textwidth} p{0.07\textwidth}} 
 \hline
 Pass Name & \multicolumn{3}{c}{Gate Count Reduction} \\ 
  \hline
  {} & \textbf{Min}  & \textbf{Max}   &  \textbf{Avg.} \\ 
rotation-folding & 0.6\% & 34.4\% & 18.2\% \\
single-qubit-gate-merging & 0.0\% & 41.3\% & 6.2\% \\
circuit-optimizer & 0.0\% & 12.9\% & 5.8\% \\
voqc & 8.2\% & 38.6\% & 22.6\% \\
 \hline
\textbf{Level 1} & 8.8\% & 42.0\% & 23.2\% \\
\end{tabular}
\end{table}

Not only does \texttt{qcor} offer an effective quantum circuit optimization solution, it also incorporates state-of-the-art qubit placement techniques, as described in Section~\ref{sec:Placement}. For near-term quantum devices with limited connectivity, efficient qubit placement is of great importance to the fidelity and success rate of circuit execution.
\begin{table}[!t]
\caption{The number of two-qubit gates after placement using various placement strategies. For each benchmark case, the best result among Sabre ($N_{sabre}$), \texttt{swap-shortest-path} ($N_{ssp}$), and QX-mapping ($N_{QX}$) is shown in boldface. $n$ is the number of qubits and $N$ is the number of two-qubit gates before placement. The improvement percentage is relative to that of \texttt{swap-shortest-path}.}
\label{tab:placement-data}
 \begin{tabular}{ p{0.09\textwidth}  p{0.03\textwidth} p{0.05\textwidth} p{0.06\textwidth} p{0.06\textwidth} p{0.06\textwidth} p{0.07\textwidth}} 
 \hline
 Name & $n$ & $N$ & $N_{ssp}$ & $N_{sabre}$ & $N_{QX}$ & Imp. [\%]  \\ 
  \hline
barenco10 & 19 & 190& 883 & \textbf{526} & 724 & 40.4\\
barenco5 & 9 & 70 & 211 & \textbf{175} & 250 & 17.1\\
grover5 & 9 & 248 & 1158 & \textbf{686} & 1100 & 40.8\\
hwb6 & 7 & 110 & 445 & \textbf{305} & 449 & 31.5\\
hwb8 & 12 & 6741 & 39988 & \textbf{22716} & 31263 & 43.2\\
mod5\_4 & 5 & 28 & 106 & \textbf{55} & 70 & 48.1\\
qft4 & 5 & 46 & 154 & 112 & \textbf{109} & 29.2\\
tof3 & 5 & 16& 55 & \textbf{31} & 40 & 43.6\\
tof5 & 9 & 36 & 171 & \textbf{108} & 144 & 36.8\\
vbe3 & 10 & 58 & 220 & \textbf{127} & 202 & 42.3\\
\end{tabular}
\end{table}
In Table~\ref{tab:placement-data}, we show a comparison in terms of gate count between some of the placement options that are available in \texttt{qcor}. In these test cases, we have processed the input circuits through \texttt{qcor} circuit optimization passes before performing hardware placement. The target device is the 65 qubit IBM \texttt{ibmq\_manhattan} backend, which has a heavy-hexagon lattice topology. We measured the improvement percentage of built-in placement strategies by comparing the number of two-qubit gates present in the benchmark circuit after placement against that of the \texttt{qcor} default Staq \texttt{swap-shortest-path}~\cite{staq} strategy. As can be seen in Table~\ref{tab:placement-data}, \texttt{qcor's} placement improves the number of added two-qubit gates from 17\% up to 48\% compared to that of Staq.

\subsection{QCOR-Enabled Library Development}
% vqe, qaoa, adapt
Finally, we turn our attention to future design goals with regards to the \texttt{qcor} compiler and runtime library. We wish to demonstrate how one might leverage the infrastructure and compiler defined in this work for high-level quantum algorithmic library development. It is our intention that the work described here will form the basis for the creation of scientific libraries that hide or abstract away the low-level machinery required for quantum-classical algorithm implementation. Specifically, here we introduce a prototype library called \texttt{qcor\_hybrid} that provides high-level data structures for common hybrid, variational quantum-classical algorithms. We demonstrate how this library enables the integration of the VQE \cite{VQE} and ADAPT \cite{adapt_vqe} algorithms within existing C\texttt{++} applications. 

\begin{figure}[b!] 
  \lstset {language=C++}
  \begin{lstlisting}
#include "qcor_hybrid.hpp"
__qpu__ void ansatz(qreg q, std::vector<double> p) 
{
  X(q[0]);
  auto exp_arg = X(0) * Y(1) - Y(0) * X(1);
  exp_i_theta(q, p[0], exp_arg);
}

int main(int argc, char **argv) {
  // Define the Hamiltonian using the QCOR API
  auto H = 5.907 - 2.1433 * X(0) * X(1) -
           2.1433 * Y(0) * Y(1) + .21829 * Z(0) -
           6.125 * Z(1);

  // Create the VQE instance, giving it the kernel,
  // the Hamiltonian, and an extra option to run 
  // each point 10 times to gather statistics
  VQE vqe(ansatz, H, 
        {{"vqe-gather-statistics",10}});
  
  // Loop over 20 points in [-1., 1.] 
  // and compute the energy at that point
  for (auto [iter, x] :
       enumerate(linspace(-1., 1., 20))) {
    std::cout << iter << ", " 
                << x << ", " 
                << vqe({x}) << "\n";
  }

  // Dump the data to file for processing
  vqe.persist_data("param_sweep_data.json");
}
-------------- compile/run with ----------------
// Exact execution
$ qcor -qpu qpp vqe.cpp && ./a.out
// Noisy execution
$ qcor -qpu aer[noise-model:custom_noise.json] 
    vqe.cpp && ./a.out
// Error mitigated execution 
// (apply readout error mitigation)
$ qcor -qpu aer[noise-model:custom_noise.json] 
    vqe.cpp -em ro-error && ./a.out
\end{lstlisting}
\caption{Code snippet demonstrating the \texttt{qcor\_hybrid} \texttt{VQE} data structure.}
\label{fig:qcor_hybrid_vqe}
\end{figure}
\subsubsection{VQE}
\texttt{qcor\_hybrid} provides a \texttt{VQE} data structure that hides the complexity of the \texttt{qcor} data model and asynchronous execution API. Programmers simply instantiate this data structure, invoke its \texttt{execute()} method, and retrieve the optimal energy and associated parameters. Moreover, one can use the data structure without the full optimization loop, and simply invoke an \texttt{operator()(std::vector<double>)} method to evaluate the expectation value of the given \texttt{Operator} at the provided parameters.  
\begin{figure}[b!] 
\includegraphics[width=.5\textwidth]{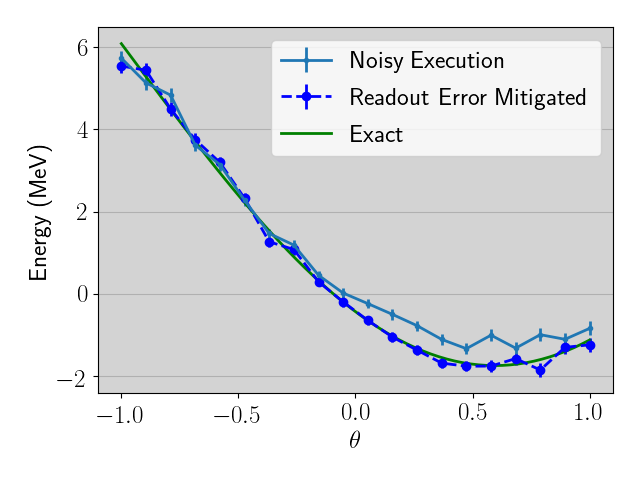}
\caption{Results of running the code in Figure \ref{fig:qcor_hybrid_vqe} with differing \texttt{qcor} command line arguments (shown at bottom of Figure \ref{fig:qcor_hybrid_vqe}).}
\label{fig:vqe_noise_plot}
\end{figure}

The code snippet in Figure \ref{fig:qcor_hybrid_vqe} demonstrates the use of \texttt{qcor\_hybrid} for sweeping the variational parameter for a prototypical state preparation circuit and computing the associated expectation value of the given \texttt{Operator}. Programmers begin by including the library header file, followed by the definition of a parameterized quantum kernel. Programmers instantiate an \texttt{Operator} representation of the Hamiltonian in the same way as previous examples. The \texttt{VQE} data structure is instantiated, taking a reference to the quantum kernel and Hamiltonian. Extra options can be provided to influence the execution, and here we demonstrate requesting that each point be computed multiple times to gather appropriate statistics. Computation of the expected value is affected via the \texttt{operator()()} method on the VQE class. At the command line, one can specify which backend this code should be compiled for. We demonstrate the compilation and execution of this code for a noise-free, exact backend, a noisy simulation backend, and a noisy simulation backend with readout-error mitigation applied. The results for these three executions are shown in Figure \ref{fig:vqe_noise_plot}.

\subsubsection{ADAPT}

%Variational quantum algorithms have found ample applications in broad range of problems, with VQE and QAOA standing out. While the former, and its many variants, is somewhat ubiquitous in physical sciences, such as quantum chemistry and quantum field theories, QAOA has enjoyed wide success in combinatorial optimization. Despite some intrinsic differences, they share the property that the optimization of the expectation value of the problem Hamiltonian relies on a predetermined form of the parameterized ansatz circuit. This constraint may raise questions about the suitability of the corresponding trial wave function to approach the desired ground state and tends to place a burden on the classical optimization, as not all parameters in the ansatz are of significance to the problem, but that knowledge is not communicated to the classical optimizers, which will treat all parameters on an equal footing.
\begin{figure}[b!] 
  \lstset {language=C++}
  \begin{lstlisting}
// QCOR hybrid algorithms library
#include "qcor_hybrid.hpp"

// Define the state preparation kernel
__qpu__ void initial_state(qreg q) {
  X(q[0]);
  X(q[1]);
  X(q[4]);
  X(q[5]);
}
int main() {
  // Define the Hamiltonian using the QCOR API
  auto H = 0.111499 * Z(0) * Z(6) + ...;

  // optimizer
  auto optimizer = createOptimizer(
      "nlopt", {{"nlopt-optimizer", "l-bfgs"}});

  // Create ADAPT-VQE instance
  ADAPT adapt(initial_state, H, optimizer,
                {{"sub-algorithm", "vqe"},
                {"pool", "singlet-adapted-uccsd"},
                {"n-electrons", 4},
                {"gradient_strategy", "central"}});

  // Execute and print
  auto energy = adapt.execute();
  std::cout << energy << "\n";
}
-------------- compile/run with ----------------
$ qcor -qpu tnqvm adapt-vqe.cpp && ./a.out
\end{lstlisting}
\caption{Code snippet demonstrating the ADAPT algorithm.}
\label{fig:qcor_adapt_example}
\end{figure}
\begin{figure}[t!] 
\includegraphics[width=.5\textwidth]{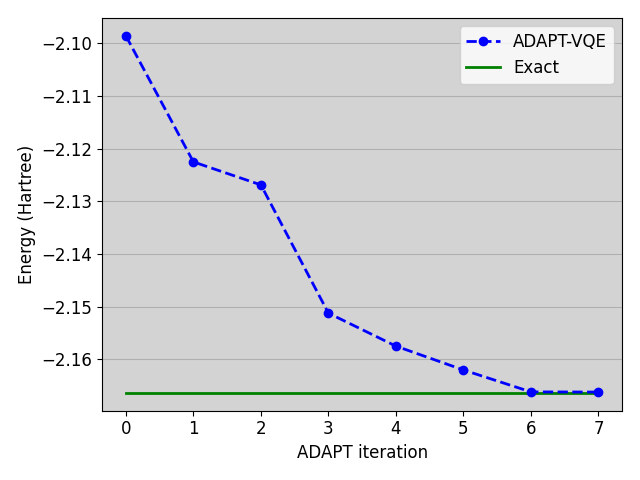}
\caption{Results of running the code in Figure \ref{fig:qcor_adapt_example} with the TNQVM as the noiseless numerical simulator.}
\label{fig:adapt_h4}
\end{figure}
The \texttt{qcor\_hybrid} library provides a high-level data structure implementing the popular ADAPT (Adaptive Derivative Assembled Problem Tailored) algorithm, which builds an adaptive circuit ansatz on-the-fly that varies according to the complexity of the problem at hand. The ADAPT algorithm provides an iterative loop that checks for the most relevant operator (Pauli or fermionic), updates the ansatz, and proceeds by calling either the VQE or QAOA routine, depending on the problem of interest. Detailed accounts on these two instances can be found elsewhere \cite{adapt_vqe, adapt_qaoa}.
The code snippet in Figure \ref{fig:qcor_adapt_example} illustrates how to instantiate and run an ADAPT-VQE simulation of %the hydrogen molecule
a chain of four hydrogen atoms taking advantage of the \texttt{qcor\_hybrid} library. This is followed by the definition of the quantum kernel representing the initial state. The \texttt{main()} function body contains the definitions for the problem Hamiltonian, shortened here for the sake of clarity, and the desired optimizer, which are followed by problem- and sub-algorithm-specific parameters. In this case, we need to pass to the ADAPT instance the variational algorithm it will employ to optimize the circuit, the number of electrons, and the set of fermionic operators associated with the variational parameters. Because the chosen optimization strategy here supports gradients (L-BFGS), to aid in updating the variational parameters, we also provide the algorithm with a strategy for its computation (numerical central finite differences). The first three arguments in the constructor of the \texttt{ADAPT} class are the necessary components shared by both VQE and QAOA, namely initial state, \texttt{Operator}, and \texttt{Optimizer}, while the last argument is an options map that is responsible for passing the problem- and sub-algorithm-specific parameters. A simulation exemplifying the code snippet in Figure \ref{fig:qcor_adapt_example} is presented in Figure  \ref{fig:adapt_h4}.

\section{Conclusion}
We have presented \texttt{qcor}, a language extension to C\texttt{++} and associated compiler executable that enables heterogeneous quantum-classical computing in a single-source C\texttt{++} context. Our approach leverages a novel domain specific language pre-processing plugin from Clang (the \texttt{SyntaxHandler}) and enables general quantum DSL integration as part of quantum kernel expression. Moreover, we build upon the XACC quantum programming framework, thereby enabling a hardware-agnostic retargetable compiler, in addition to an integration mechanism for common quantum compiling, optimization, and qubit placement tasks. We believe that \texttt{qcor} will ultimately promote tight integration of future quantum co-processors with existing high-performance computing application software stacks. Finally, we note that our work is completely open source and available at \url{https://github.com/ornl-qci/qcor}. 

\section*{Acknowledgment}
This work has been supported by the US Department of Energy (DOE) Office of Science Advanced Scientific Computing Research (ASCR) Quantum Computing Application Teams (QCAT), Quantum Testbed Pathfinder (QTP), and Accelerated Research in Quantum Computing (ARQC).  This work was
also supported by the ORNL Undergraduate Research Participation Program, which is sponsored by ORNL and administered jointly by ORNL and the Oak Ridge Institute for Science and Education (ORISE). ORNL is managed by UT-Battelle, LLC, for the US Department of Energy under contract no. DE-AC05-00OR22725. This research used resources of the Oak Ridge Leadership Computing Facility at the Oak Ridge National Laboratory, which is supported by the Office of Science of the U.S. Department of Energy under Contract No. DE-AC05-00OR22725. This research used resources of the Argonne Leadership Computing Facility, which is a DOE Office of Science User Facility supported under Contract DE-AC02-06CH11357.

\bibliographystyle{plain}
\bibliography{main}
\end{document}